\documentclass[10pt]{article} 

\usepackage[utf8]{inputenc}
\usepackage{amsmath}
\usepackage{amsfonts}
\usepackage{amssymb}
\usepackage{mathtools}
\usepackage{listings}
\usepackage{mathrsfs}
\usepackage{times}
\usepackage{float}
\usepackage{color}
\usepackage{setspace}
\usepackage{color,soul}

\usepackage{amsmath,amsfonts,amsthm,eucal}
\usepackage{enumerate}
\usepackage[letterpaper,margin=3cm]{geometry}
\usepackage{float}
\usepackage[title]{appendix}
\usepackage{color}
\usepackage{tabularx}
\usepackage{siunitx}
\usepackage[tableposition=below]{caption}
\usepackage{array}

\usepackage[
pdfstartview=XYZ,
bookmarks=true,
colorlinks=true,
linkcolor=blue,
urlcolor=blue,
citecolor=blue,
pdftex,
bookmarks=true,
linktocpage=true,   
hyperindex=true
]{hyperref}

\usepackage{lipsum}
\usepackage{lineno}

\usepackage[FIGBOTCAP,TABTOPCAP,bf,tight]{subfigure}

\usepackage{natbib}
\setcitestyle{authoryear}

\usepackage[pdftex]{graphicx}
\usepackage{epstopdf}
\usepackage{booktabs} 
\usepackage{tikz,mathpazo}
\usetikzlibrary{shapes.geometric, arrows}
\usepackage{caption}

\newcommand{\tensor}[1]{\ensuremath{\boldsymbol{#1}}}

\usepackage{algorithm}
\usepackage{algorithmicx}
\usepackage[noend]{algpseudocode}

\theoremstyle{remark}

\theoremstyle{definition}

\usepackage{algorithm}
\usepackage[noend]{algpseudocode}
\newcolumntype{M}[1]{>{\centering\arraybackslash}m{#1}}

\title{Synthesizing realistic sand assemblies with denoising diffusion in latent space} 

\begin{document}

\author{
    Nikolaos N. Vlassis\thanks{Corresponding author, Department of Mechanical and Aerospace Engineering, Rutgers University, Piscataway, NJ 08854. \textit{nick.vlassis@rutgers.edu}} \and
    WaiChing Sun\thanks{Department of Civil Engineering and Engineering Mechanics, Columbia University, New York, NY 10027. \textit{wsun@columbia.edu}} \and
    Khalid A. Alshibli\thanks{Department of Civil and Environmental Engineering, University of Tennessee, Knoxville, TN 37996. \textit{alshibli@utk.edu}} \and
    Richard A. Regueiro\thanks{Department of Civil, Environmental, and Architectural Engineering, University of Colorado, Boulder, CO 80309. \textit{richard.regueiro@colorado.edu}}
}

\maketitle

\begin{abstract}
The shapes and morphological features of grains in sand assemblies have far-reaching implications in many engineering applications, such as geotechnical engineering, 
computer animations, petroleum engineering, and concentrated solar power.  
Yet,  our understanding of the influence of grain geometries on macroscopic response is often only qualitative,  due to the limited availability of high-quality 3D grain geometry data. 
In this paper, we introduce a denoising diffusion algorithm
that uses a set of point clouds collected from the surface of individual sand grains to generate grains in the latent space. 
By employing a point cloud autoencoder, the three-dimensional point cloud structures of sand grains are first encoded into a lower-dimensional latent space. 
A generative denoising diffusion probabilistic model is trained to produce synthetic sand
that maximizes the log-likelihood of the generated samples belonging to the original data distribution measured by a Kullback-Leibler divergence. 
Numerical experiments suggest that the proposed method is capable of generating realistic grains with morphology, shapes and sizes 
consistent with the training data inferred from an F50 sand database . 
We then use a rigid contact dynamic simulator to pour the synthetic sand 
in a confined volume to form granular assemblies in a static equilibrium state with targeted distribution properties.
To ensure third-party validation, 50,000 synthetic sand grains and the 1,542 real synchrotron microcomputed tomography (SMT) scans of the F50 sand,  as well as the granular assemblies composed of synthetic sand grains are made available in an open-source repository.
\end{abstract}

\section{Introduction}
\label{intro}

Sand is the second most consumed natural resource material in the world \citep{beiser2019world} (the first being water). 
Every year, 50 billion tonnes of sand are used as raw material for concrete, glass, and electronics. 
Significant progress has been made on understanding the macroscopic mechanical behavior of sand over the last several decades, noticeably in phenomenological modeling under confining pressure \citep{nova1979constitutive, mitchell2005fundamentals, jefferies1993nor, dafalias2004simple} and granular flow rheology \citep{wieghardt1975experiments, kamrin2019non, moriguchi2009estimating}. However, many of those studies often focus on extensions of granular assemblies composed of grains of simpler shapes such as spherical glass beads and ellipsoids \citep{lin1997three, kuhn2004contact, peters2005characterization}.

Meanwhile, recent work on both micro-computed-tomography (CT) experiments and dynamic imaging analysis has revealed that grain shapes 
may have profound effects on the onset of shear banding as well as shear strength of granular materials \citep{kawamoto2018all, altuhafi2016effect}, although the exact role or mechanism of shape effects on natural sand assemblies has yet to reach a complete consensus in the literature. 
Another promising research direction to gain quantitative understanding of shape effects on maroscopic sand behavior is to introduce algorithms to create 
realistic shapes of sand grains, both for the sake of generating numerical specimens for simulations (e.g., \citet{cil2015modeling, duriez2021precision}), 3D printing of grains for repeatable control experiments (e.g., \citet{gupta2019open, ahmed2021triaxial}),
analyzing via shape descriptors (e.g. \citet{bowman2001particle}), and uncertainty quantification. 
Popular algorithms used to generate synthetic shapes of sand grains 
included Fourier shape descriptors \citep{thomas1995use, bowman2001particle, mollon2013generating, o2017relating} and genetic algorithms \citep{jerves2017geometry, buarque2018granular}. 
In the former case, shape features collected from 3D scans of sand grains can be obtained via discrete Fourier Transform. 
Synthetic sand can then be reconstructed by matching the occurrence statistics of the Fourier shape features of those obtained from the real sand grains. 
The upshot of the Fourier descriptors is that they are inherently translation invariant. 
Furthermore,  the resolution of the grain geometry can be controlled by filtering the high-frequency components of the Fourier Series. 
However,  Fourier transform does not allow sharp or locally refined features, as well as broken or mixed shapes to be easily reproduced \citep{nixon2019feature}. 
\citet{jerves2017geometry}, meanwhile, 
 considered the Fourier series and spherical haromonics approach as a mathematical and numerical way of understanding the grain morphology, and propose the usage of morphological DNA to study the real grains via more conventional descriptors, such as aspect ratio, roundness and principal geometric directions. 
 This idea simplifies the procedure to generate grain shapes, but the perceptions of correct reconstruction are limited to what can be described by the chosen shape descriptors. 
 Unlike the Fourier descriptors which guarantee the convergence of grain shape, the clone approach cannot exert controls on the shape features that are not covered by the set of the shape descriptors used as criteria to generate the grain \ avatar. 

In this paper, our goal is to introduce a new alternative algorithm where there are no hand-picked descriptors used to reconstruct the grains. Instead, a new class of generative machine learning models, denoising diffusion probabilistic models (DDPM) is used to generate highly realistic grains without the need to explicitly enforce descriptors. 
Originating from the work of \citet{sohl2015deep},  these belong to a class of generative models that learn a given data distribution by learning to reverse a gradual, multi-step noising process.  DDPMs are a further advancement of this concept,  suggested by \citet{ho2020denoising}, and established an equivalence with score-based generative models. 
These models learn a gradient of the log-density of the data distribution utilizing denoising score matching \citep{hyvarinen2005estimation}.
DDPMs have proven effective in generating highly realistic data, with applications extending to image generation \citep{song2020denoising,ramesh2022hierarchical,rombach2022high}, audio synthesis \citep{chen2020wavegrad,kong2020diffwave}, and time-series forecasting \citep{rasul2021autoregressive}. 
Our work utilizes this capacity of DDPMs to produce synthetic sand grains but not directly on the original 3D space but on its corresponding latent representation, similar to \citep{rombach2022high}.

Thus, this generation process does not require the use of any hand-picked or designated descriptors to generate new sand grains. 
Instead, a point cloud autoencoder is used to generate a pair of contracting and expansive path that maps the high-dimensional data into a low-dimensional feature space where the diffusion process takes place. 
This generic approach is highly flexible as the machine learning can generate different features tailored for the given data set, and hence bypass the usage of shape descriptors in the grain generation algorithm. As shown in the paper, this general appraoch may provide highly realistic data with comparable behaviors and statistics of size and shape measured by the hand-crafted descriptors even though there is no ad-hoc manipulation, such as genetic programming, to alter the results.

The remainder of this paper is structured as follows. In Section~\ref{sec:denoising_diffusion}, we delve into the formation of sand grain embeddings using a point cloud autoencoder, and the process of applying a denoising diffusion probabilistic model on these embeddings. 
Section~\ref{sec:sand_grain_dataset} provides insight into the F50 sand grain database used for training both the autoencoder and the denoising diffusion algorithm, with a brief overview of the 3D sand grain data acquisition methodology, statistical analysis of the grain samples, and a description of preprocessing methods employed.
Section~\ref{sec:denoising_diffusion_results} elucidates the architecture, training, and results of our point cloud convolutional autoencoder and denoising diffusion algorithm, presenting a comparison between generated and database sand grain embeddings and grains.
In Section~\ref{sec:granular_assemblies}, we illustrate our application of the denoising diffusion algorithm for generating diverse sand grain assemblies, with a special focus on adjusting generated grain attributes to construct assemblies that exhibit targeted property distributions.
Finally, Section~\ref{sec:conclucion} offers concluding remarks, encapsulating the key findings and potential future directions of our study.

\section{Denoising diffusion in latent space for point clouds}
\label{sec:denoising_diffusion}
In this section,  we discuss the point cloud representation of the sand grains, the autoencoder architecture to create latent space embeddings of said point clouds, and the DDPM model used to generate realistic sand grain embeddings.  The decision to use a point cloud representation stems from a comparison of several viable options, such as level set methods, voxel representations,  and graph-based models, among others. Point cloud representation offers a number of distinct advantages over these alternatives.

While voxel-based methods \citep{hall2010discrete,varfolomeev2019application,henkes2022three} involve intuitive representations of granular matter in a 3D space,  point clouds handle spatial data more efficiently. 
Voxel representations often result in data that are both sparse,high-dimensional and resolution-dependent, leading to computational challenges which can be a roadblock in data-driven applications. 
Point clouds, on the other hand, are adept at representing complex 3D structures in a much more compact form, making them less resource-intensive \citep{guo2020deep}.

Level set methods \citep{vlahinic2017computed,kawamoto2018all,zhang2022image} have been used to represent grain shapes using a mathematical function, typically involving a higher-dimensional space. 
While these methods are powerful in handling changes in the shape topology (like merging or splitting parts), they often require complex mathematical constructs and computations. 
Also, approximating the intricate and diverse shapes of sand grains accurately with mathematical functions can be challenging and may not always yield the most realistic representations.
These level set representations could, however, be approximated with a neural network similar to \cite{vlassis2021sobolev,vlassis2022component} which will be considered in future work.

Although an undirected graph representation of the point cloud is plausible, where the connections could be established through nearest neighbors or through a point cloud triangulation surface mesh (which we detail in later sections), we opted for a different approach. 
The graph representation undoubtedly holds advantages in encapsulating proximity or connectivity information among the points on the surface, which could be leveraged to enrich the embedding process \citep{kipf2016variational,vlassis2020geometric,vlassis2023geometric}. However, the reconstruction of point connectivity can be a complex and computationally intensive procedure \citep{menon2011link,zhang2018link}.
Therefore, our choice was to use the point cloud coordinate representation, which retains the straightforward and efficient characteristics of point clouds, while still maintaining some level of adjacency information in the embedding algorithm. 
We accomplish this by enhancing the node embeddings of the point cloud representation, the methodology of which we elaborate upon in the subsequent sections.

Point cloud models provide a simple representation of geometries that offer greater flexibility. 
By considering a set of points in space as a representation of probability distribution,  a point cloud model does not require the reconstruction of topological structures or the generation of the implicit functions \citep{williams2022neural} or manifolds \citep{williams2019deep}  to represent the 3D geometries.
They provide a direct and intuitive representation of 3D objects that aligns well with many data acquisition methods, like 3D scanning or photogrammetry, which are common in the study of sand grains.
The representation of objects is more direct and intuitive, allowing for the accurate capture and expression of the broad range of complex shapes natural sand grain structures. 
The point cloud model's ability to adjust and accommodate to these variations renders it a more suitable choice for our work.

\subsection{Point cloud autoencoder}
\label{sec:autoencoder}

In this section, we review key advancements in the application of autoencoders to point cloud data.
We also provide definitions for the point cloud representation used in this paper as well as describe our schematic for constructing a sand grain point cloud latent space that will be the basis of our diffusion algorithm.

\subsubsection{Literature review of point cloud autoencoders}
\label{sec:autoencode_lit}

Comprehensive surveys \citep{bronstein2017geometric, griffiths2019review, liu2019deep, hooda2022survey} have illustrated the potential of neural networks in handling non-Euclidean spaces and point clouds, particularly their ability to manage point sparsity and shape irregularities and capacity to tackle feature extraction, compression, segmentation, reconstruction, and generation tasks. 

For feature extraction and segmentation, \citet{elbaz20173d} demonstrated the potential of super-points and autoencoders for accurate point cloud registration, tracking, and object pose estimation. \citet{liu2019l2g} and \citet{wang2019dynamic} introduced autoencoders that not only capture local details but also the global structure of point clouds, resulting in significant performance improvements in shape classification, retrieval, and feature extraction tasks. \citet{rios2020feature} proposed a visualization method to understand the features learned by these point cloud autoencoders.

On the front of reconstruction and generation,  \citet{yang2018foldingnet} introduced FoldingNet, an end-to-end deep auto-encoder achieving low reconstruction errors and good classification performance on point clouds. \citet{mandikal20183d} proposed a latent embedding matching approach, 3D-LMNet, that reconstructs 3D point clouds from a single image.  \citet{achlioptas2018learning} made a comparison between Generative Adversarial Networks (GANs) and Gaussian Mixture Models (GMMs) trained in the latent space of autoencoders. 
The recent work by \citet{gadelha2018multiresolution} introduced multi-resolution tree-structured networks, showing the potential for efficient processing of point clouds, while \citet{wiesmann2021deep} presented a deep convolutional autoencoder architecture that demonstrated improved reconstructions at the same bit rate and improving on point cloud compression. 

In our work, we adapt a 1D convolutional autoencoder architecture similar to \citet{achlioptas2018learning} and modify it to construct a latent space suitable for our diffusion algorithm to learn.
The primary purpose of this autoencoder is to learn a compact and efficient representation of the grains' structure and geometry in a lower-dimensional space, which will later be useful for generating new grains. 
The autoencoder is composed of two key parts: an encoder and a decoder. 
The role of the encoder is to take the original point cloud data and compress it into a lower-dimensional latent space. 
The decoder, on the other hand, uses these latent space embeddings to reconstruct the original point cloud data.

\subsubsection{Point cloud autoencoder for latent space representation}
\label{sec:autoencoder_theory}

A sand grain in this work is represented by a point cloud in three-dimensional Euclidean space $\mathbb{R}^3$. 
Each point $p_i$ in the point cloud can be defined as an ordered triple $(x_i, y_i, z_i)$, where $x_i, y_i, z_i \in \mathbb{R}$. 
Hence, a point cloud $\tensor{P}$ can be defined as:
\begin{equation}
\tensor{P} = \{p_1, p_2, p_3, \ldots, p_N\} \subset \mathbb{R}^3.
\label{eq:pointcloud}
\end{equation}
Here, each $p_i = (x_i, y_i, z_i)$ and $i \in {1, 2, \ldots, N}$. 
Therefore, $\tensor{P} $ is a finite set of $N$ points in $\mathbb{R}^3$.

Each point in the point cloud can also be associated with additional features $\tensor{\phi}$. 
For instance, each point $p_i$ might have $F$ extra features, that hold information about point adjacency, surface normals, and surface curvature among others. 
This enhanced point cloud can be represented as a matrix $\overline{\tensor{X}}$ of $N \times (3+F)$ dimensions, where $N$ is the number of points and $F$ is the total number of features per point including the 3D Cartesian coordinates. 
Each row in this matrix represents a point, and each column represents a feature such that $\overline{X}_{ij}$ denotes the $j$-th attribute of the $i$-th point, for $i \in {1, 2, \ldots, N}$ and $j \in {1, 2, \ldots, 3+F}$. The first three attributes correspond to the $x, y, z$ coordinates, and the remaining $F$ attributes correspond to other features $\phi_k$ of the point for $k \in {1, 2, \ldots, F}$. 
We also define the simpler spatial representation $\widehat{\tensor{X}}$ of the point cloud that contains only the spatial coordinates of the points and is sufficient to represent the 3D structure. 

The autoencoder architecture $\tensor{\zeta}$ in this work maps the enhanced point cloud representation $\overline{\tensor{X}}$ to the spatial representation of the point cloud $\widehat{\tensor{X}}$.
We define the encoder component of the architecture as a neural network mapping of $\overline{\tensor{X}}$ to an latent space embedding $\tensor{x}$, parameterized by weights $\tensor{W}_\text{e}$ and biases $\tensor{b}_\text{e}$.
The decoder architecture is a neural network mapping from the embedding $\tensor{x}$ to the point cloud $\widehat{\tensor{X}}$ parameterized by weights $\tensor{W}_\text{d}$ and biases $\tensor{b}_\text{d}$.
It is noted that we incorporate the enhanced features in the point cloud representation $\overline{\tensor{X}}$ to achieve augmented results in the embedding process.
The decoder only outputs the spatial representation of the point cloud to simplify the learning problem of the decoder.
Additional information of the reconstructed point cloud's surface representation can be obtained in post-processing but is not necessary to describe the 3D grain shape.

The autoencoder is trained and validated utilizing the Chamfer Distance 3D loss function as the objective criterion. This choice of loss function enables the autoencoder to effectively learn the underlying structure of the point cloud data by minimizing the point-wise distance between the input and the reconstructed output. 
The Chamfer loss is specifically tailored for point cloud data \citep{achlioptas2018learning,urbach2020dpdist,wu2021density} due to its computational efficiency,  point order-invariance, and robustness to differences in point density.  
It calculates the sum of the distances from each point in the input point cloud $\tensor{P}_A$ to its nearest neighbor in the reconstructed point cloud $\tensor{P}_B$, and vice versa. 
The loss is defined as:
\begin{equation}
L_{CD}(\tensor{P}_A, \tensor{P}_B) = \frac{1}{|\tensor{P}_A|} \sum_{p_A \in \tensor{P}_A} \min_{p_B \in \tensor{P}_B} ||p_A - p_B||^2 + \frac{1}{|\tensor{P}_B|} \sum_{p_B \in \tensor{P}_B} \min_{p_A \in \tensor{P}_A} ||p_A - p_B||^2.
\label{eq:chamfer}
\end{equation}
Here, $||p_A - p_B||^2$  represents the squared Euclidean distance between points $p_A$ and $p_B$. By minimizing this loss, the autoencoder learns to generate a reconstruction that closely approximates the original point cloud in terms of both structure and geometry. 
In general, the Chamfer loss normalizes the calculated distance by the number of points in the input point cloud ($|\tensor{P}_A|$) and the reconstructed point cloud ($|\tensor{P}_B|$), making the comparison between point clouds of varying sizes more meaningful. 

Thus, we can define the autoencoder training and latent space embedding objective, modeled after the Chamfer Distance 3D loss function, for the point cloud $\tensor{P}_m$ training samples for $m \in {1, 2, \ldots, M}$ in the database as:
\begin{equation}
\tensor{W}^{\prime}_{\text{e}}, \tensor{b}^{\prime}_{\text{e}}, \tensor{W}^{\prime}_{\text{d}}, \tensor{b}^{\prime}_{\text{d}}= \underset{\tensor{W}_{\text{e}}, \tensor{b}_{\text{e}}, \tensor{W}_{\text{d}}, \tensor{b}_{\text{d}}}{\operatorname{argmin}} \left(\frac{1}{M} \sum_{m=1}^{M} \left[ L_{CD}\left(\tensor{P}_m, \tensor{\zeta}\left(\tensor{P}_m\right)\right) \right]\right).
\label{eq:embedding_objective}
\end{equation}
In the above equation,  $\tensor{\zeta}({\tensor{P}}_m)$ represents the reconstructed point cloud for the $m$-th training sample.
After the optimization of the autoencoder, we can generate $\tensor{x}_m$ embeddings for each sand grain point cloud of the database. 
The specifics of the 1D convolutional point cloud autoencoder architecture and training are presented in Section~\ref{sec:autoencoder_training}.

\subsection{Denoising diffusion algorithm for realistic sand grain generation}
\label{sec:denoising_architecture}

\begin{figure}[h!]
\centering
\includegraphics[width=.85\textwidth ,angle=0]{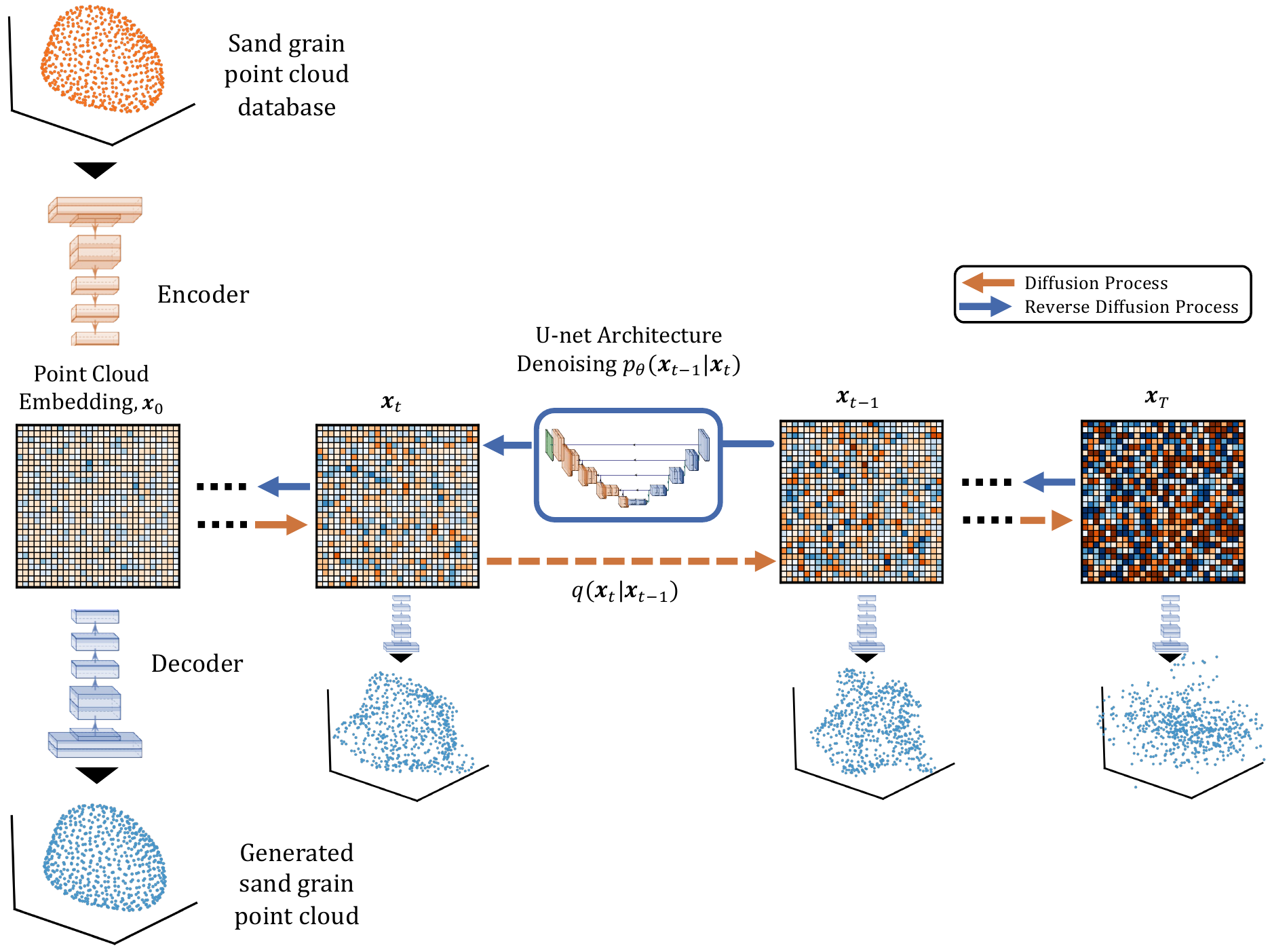} 
\caption{Schematic of the sand grain generation algorithm. The sand grain point clouds are embedded in a latent dimension with a point cloud autoencoder architecture.
Noise is progressively added to the point cloud embeddings to completely destroy the signal (diffusion process $q$). 
The denoising process $p_\theta$ is learned by a U-net neural network. 
Starting from random noise and denoising progressively,  a realistic embedding is generated and decoded to construct the generated sand grain point cloud.}
\label{fig:diffusion_process}
\end{figure}

Figure~\ref{fig:diffusion_process} depicts our schematic for creating sand grain point clouds using DDPMs in latent space representations. 
The DDPM adds Gaussian noise into the point cloud embeddings gradually,  similar to manipulating color channels in image synthesis \citep{li2018transfer}, until the signal is completely destroyed. 
This forward noising/diffusion process can be described as a Markov chain process, wherein each transition's probability is influenced solely by the present state and the time lapsed, irrespective of the preceding sequence of states.
Markov chain modeling offers simplicity, scalability, and versatility, facilitating step-wise noise addition in diffusion probabilistic models by relying only on the current state, not the entire history, thus easing the learning and generation processes.

This work follows the formulation of the forward and reverse diffusion process introduced in \cite{ho2020denoising} and \cite{nichol2021improved}.
We consider a data distribution $x_0 \sim q(x_0)$, where $x_0$ is the latent space representation of the sand grain point clouds as obtained from the autoencoder described in the previous section.
We progressively add Gaussian noise through the forward noising process $q$ to produce noised samples $x_1,...,x_T$ for $T$ diffusion time steps.
The noising is a Markov process where every noised sample $x_t$ is given by:
\begin{equation}
q(x_t | x_{t-1}) := \mathcal{N}(x_t; \sqrt{1-\beta_t} x_{t-1}, \beta_t \tensor{I}),
\label{eq:noising}
\end{equation}
where $\beta_t \in (0,1)$ is a variance schedule. 
Using Bayes theorem,   \cite{ho2020denoising} find the posterior to connect two arbitrary steps:
\begin{equation}
q(x_{t-1}|x_t,x_0) = \mathcal{N}(x_{t-1}; \tilde{\mu}(x_t, x_0), \tilde{\beta}_t \tensor{I}).
\label{eq:posterior}
\end{equation}
The mean of the Gaussian is:
\begin{equation}
\tilde{\mu}_t(x_t,x_0) := \frac{\sqrt{\bar{\alpha}_{t-1}}\beta_t}{1-\bar{\alpha}_t}x_0 + \frac{\sqrt{\alpha_t}(1-\bar{\alpha}_{t-1})}{1-\bar{\alpha}_t} x_t,
\label{eq:mu}
\end{equation}
and the variance is:
\begin{equation}
\tilde{\beta}_t := \frac{1-\bar{\alpha}_{t-1}}{1-\bar{\alpha}_t} \beta_t,
\label{eq:var}
\end{equation}
where $\alpha_{t} := 1-\beta_t$ and $\bar{\alpha}_t:=\prod_{s=0}^{t}\alpha_s$.

In this work, we focus on unconditional diffusion for generating sand.
This technique synthesizes sand grains without imposing constraints and guidance to mirror the distribution of properties in the database. 
For the scope of this work, we aim to massively generate sand grains from the dataset distribution and sub-sample the generated grains to create assemblies with targeted properties (Section~\ref{sec:granular_assemblies}).
For the use of denoising diffusion for the generation of microstructures with targeted material behaviors the reader is referred to \cite{vlassis2023denoising}.

The DDPM in this work aims to reverse the noising process by learning 
the reverse distribution $q(x_{t-1}|x_t)$.
We can then sample backwards from step $t = T$ to $t=0$ to gradually remove the noise and generate a latent space embedding $\tensor{x}$ that can be decoded into a realistic sand grain. 
When $T\to \infty$ and $\beta_t \to 0$,  $q(x_{t-1}|x_t)$ approaches a diagonal Gaussian distribution.  A network can learn the mean $\mu_\theta$ and the diagonal covariance $\Sigma_\theta$ \citep{sohl2015deep} such that:
\begin{equation}
p_{\theta}(x_{t-1}|x_t) := \mathcal{N}(x_{t-1}; \mu_{\theta}(x_t, t), \Sigma_{\theta}(x_t, t)).
\label{eq:p_theta}
\end{equation}

Following \cite{nichol2021improved}, we parameterize both $\mu_\theta$ and $\Sigma_\theta$ as this was observed to achieve better log-likelihoods. 
We define a network to learn $\epsilon_\theta(x_t,t)$ such that:
\begin{equation}
\mu_{\theta}(x_t, t) = \frac{1}{\sqrt{\alpha_t}} \left( x_t - \frac{\beta_t}{\sqrt{1-\bar{\alpha}_t}} \epsilon_{\theta}(x_t, t) \right).
\label{eq:mu_theta}
\end{equation}
Sample $x_t$, the model also outputs a vector $v$ to learn $\Sigma_\theta(x_t,t)$ such that:
\begin{equation}
\Sigma_{\theta}(x_t, t) = \exp(v \log \beta_t + (1-v) \log \tilde{\beta}_t).
\label{eq:sigma}
\end{equation}

Thus,  the forward noising process $q$ can be seen as a encoder while $p_\theta$ can be considered a decoder parameterized as a neural network -- in this case a U-net model. Together they constitute a variational autoencoder similar to \cite{kingma2013auto} that will be optimized for a variational lower bound for $p_\theta$.
The training of the DDPM optimizes two objectives simultaneously to learn both $\mu_\theta$ and $\Sigma_\theta$.
The first learning objective optimizes $\mu_\theta$ using the expectation:
 \begin{equation}
L_{\mu} = E_{t,x_0,\epsilon}\left[ || \epsilon - \epsilon_{\theta}(x_t, t) ||^2 \right],
\label{eq:L_mu}
\end{equation}
by randomly sampling $t$ to estimate a variational lower bound.
The second learning objective is used to only optimize the learned $\Sigma_{\theta}$ using:
 \begin{equation}
 L_{\text{vlb}} := L_0 + L_1 + ... + L_{T-1} + L_T, 
 \label{eq:Lvlb}
\end{equation}
where:
\begin{equation}
L_t := 
\begin{cases}
    -\log p_{\theta}(x_0 | x_1) & \text{if } t=0 \\
    D_{KL}\left( q(x_{t-1}|x_t,x_0)\|p_{\theta}(x_{t-1}|x_t) \right)& \text{if } 0 < t < T \\
    D_{KL}\left(q(x_T | x_0)\|p(x_T)\right)& \text{if } t =T
\end{cases}.
\label{Ls}
\end{equation}
Each term in the above optimizes the Kullback-Leibler (KL) divergence between the Gaussian distributions of two successive denoising steps.  
The learning objective for both $\mu_\theta$ and $\Sigma_\theta$ is defined as the hybrid objective:
\begin{equation}
D_{K L}(p_{\theta} \| q)=\int_x p_{\theta}(x) \log \frac{p_{\theta}(x)}{q(x)}.
\label{Ls}
\end{equation}
 The learning objective to learn both $\mu_\theta$ and $\Sigma_\theta$ is defined as the hybrid objective:
\begin{equation}
L_{\text{hybrid}} = L_{\mu} + \lambda L_{\text{vlb}},
\end{equation}
where $\lambda = 0.001$ is a scaling factor chosen as in \cite{nichol2021improved} to weigh the two objectives.

After the training of the DDPM is complete, we can start from random Gaussian noise to perform unconditional generation of realistic point cloud embeddings.
These embeddings can be decoded back to the original point cloud 3D space as illustrated in Fig.~\ref{fig:diffusion_process} for every step of the reverse of the diffusion process using the decoder architecture of Section~\ref{sec:autoencoder}.
We can thus circumvent developing a point cloud specific denoising diffusion algorithm by working on the embedded 2D latent space.

\section{F50 sand grain database}
\label{sec:sand_grain_dataset}

In this section,  we provide information on the F50 sand database used for the training of autoencoder and consequently the denoising diffusion algorithm.
The methodology for acquiring the sand grain 3D data is briefly discussed, and the statistical information of the grain samples and size of the data set is presented.
Finally, we describe the preprocessing methods used to prepare the point cloud sand grain samples for the neural network training.

\subsection{Sand grain database}
\label{sec:point_cloud_data}

\begin{figure}[h!]
\centering
\includegraphics[width=.25\textwidth, angle=0]{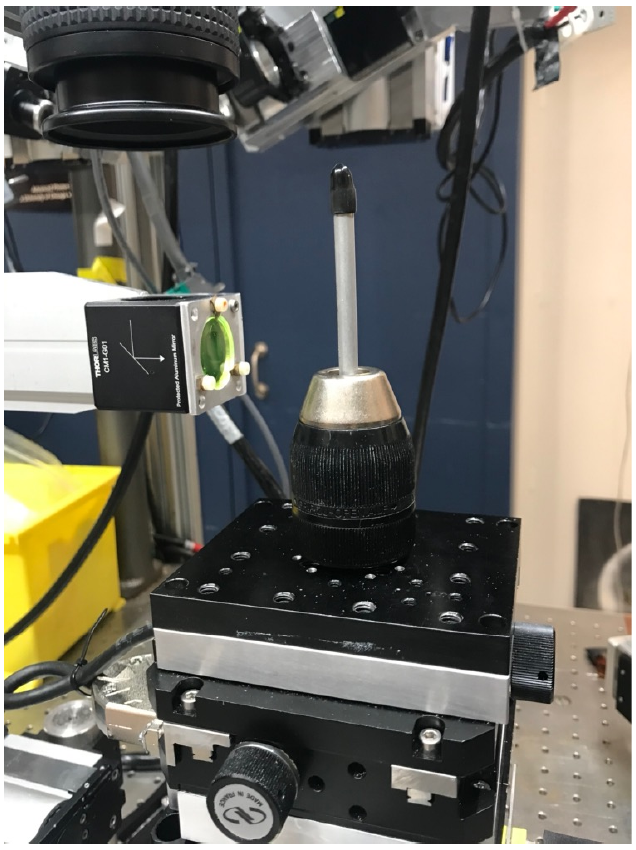} 
\caption{The Synchrotron Micro-computed Tomography (SMT) experimental setup used to acquire the F50 sand grain database.}
\label{fig:experiment_setup}
\end{figure}

The F50 sand, a natural silica sand sourced from US Silica company, is the primary constituent of the specimen used in this study. This sand is primarily composed of 99\% quartz, with grain sizes ranging between U.S. sieve \#40 (0.42 mm) and \#50 (0.297 mm). These grains were used to prepare a specimen within an aluminum tube featuring an inner diameter of 3.6 mm and an outer diameter of 6.35 mm (0.25 in.).
The specimen's 3D images were captured using the Synchrotron Micro-computed Tomography (SMT) technique at the Advanced Photon Source (APS), Argonne National Laboratory (ANL), Illinois, USA. The sand tube was mounted on the stage of beamline 13D, and fast scans were conducted using pink beam at 0.2$^\circ$ rotations, recording images while the stage rotated from 0$^\circ$ to 180$^\circ$. Each scan took an exposure time of 0.025 seconds, with an energy level ranging from 35 keV to 50 keV. A full set of SMT scans was acquired within approximately 130 seconds. The experimental setup used to acquired the database scans is shown in Fig.~\ref{fig:experiment_setup}.

\begin{figure}[h!]
\centering
\includegraphics[width=.75\textwidth ,angle=0]{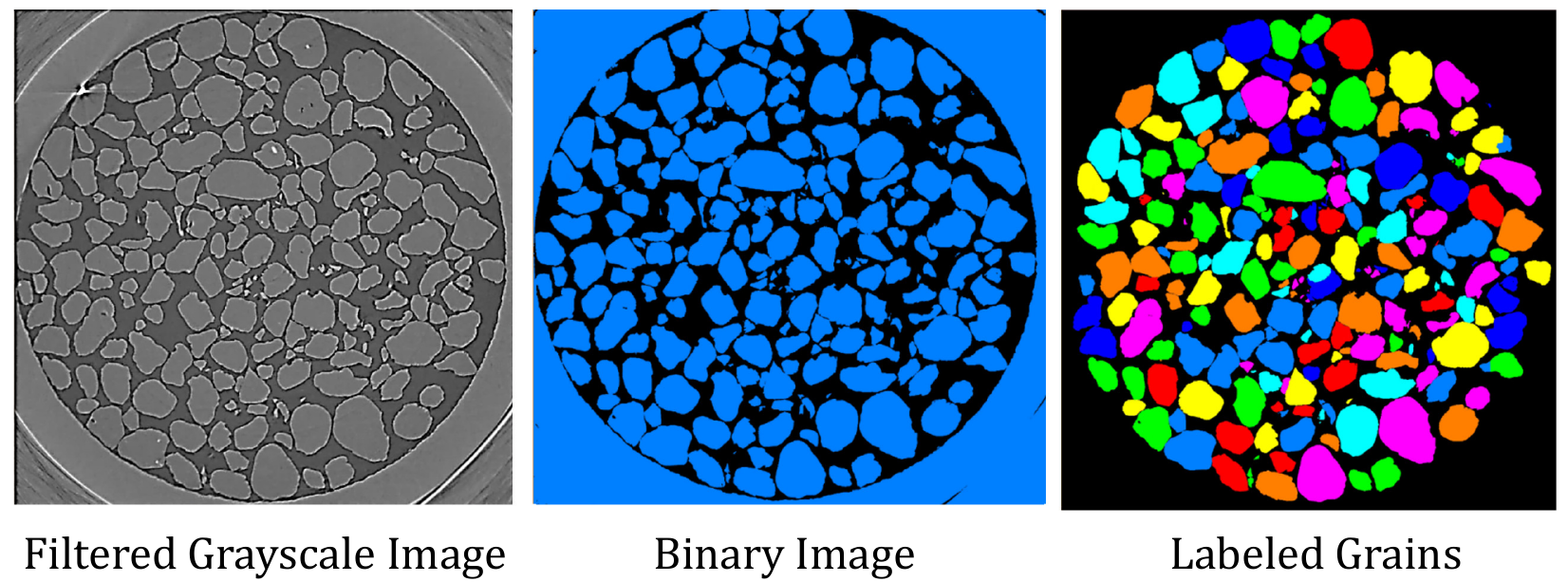} 
\caption{The post-processing procedure of labeling individual F50 sand grains from the SMT scan.}
\label{fig:image_processing}
\end{figure}

The resulting images were of dimensions 1920 x 1920 x 3434 pixels, consisting of three stacks, each with a height of 1200 pixels and with overlaps between the stacks. Each pixel corresponded to a spatial resolution of 2.06 $\mu m$. Following the image acquisition, the 3D SMT images underwent post-processing, wherein the grayscale SMT images were transformed into binary images through a thresholding process. Each grain was subsequently identified and labeled following the procedure presented in \cite{druckrey20163d}.  The post-processing of the granular assembly from the SMT scan to the identification and labeling of individual grains is illustrated in Fig.~\ref{fig:image_processing}.
The F50 sand database of 1,551 grains scanned and processed in 3D surface meshes is demonstrated in Fig.~\ref{fig:database_distributions}.

\subsection{Grain metrics}
\label{sec:grain_metrics}

In this section, we introduce the metrics employed in our study to quantitatively assess the size and shape distributions of the grains. 
Although there are numerous methods and tools available for examining size and shape distributions, we find the selected metrics to be efficient and suitable for comparing real and generated grains and assemblies within the context of this work. 
They were chosen as they offer a balance between simplicity and effectiveness to capture important characteristics without overcomplicating the evaluation process. 
While we acknowledge that this may not represent the most exhaustive method,  it was deemed an adequate and comprehensive comparison matrix.
Further exploration of additional metrics could be beneficial in future research.

\begin{figure}[h!]
\centering
\includegraphics[width=.95\textwidth ,angle=0]{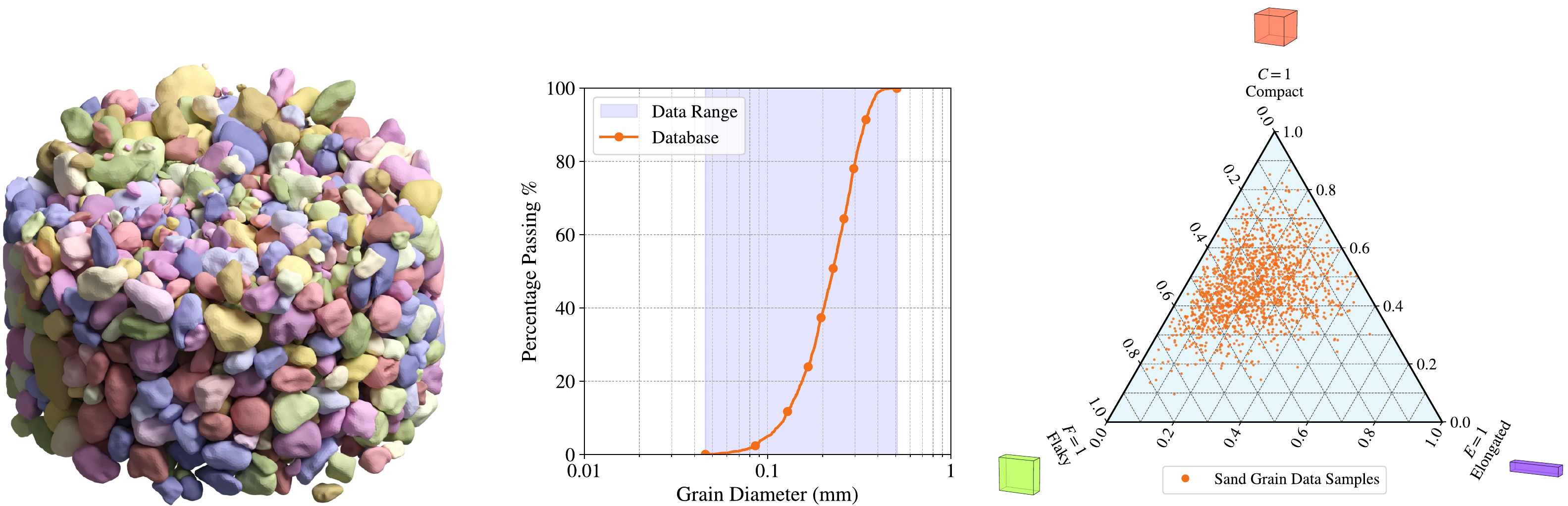} 
\caption{F50 sand database of 1,551 grains used for the sand grain generation.
The grain size distribution and grain shape metrics \citep{orosz2021surface} are also shown.}
\label{fig:database_distributions}
\end{figure}

To obtain an effective measure of grain size, we find the minimum bounding sphere for each grain. 
This method, implemented using the Trimesh library \citep{trimesh}, involves evaluating the radius of the smallest sphere that can fully enclose each point cloud representing a grain. 
The calculated diameter will be used in all of the grain size distribution curves in this work. 
This approach offers a comprehensive matrix to compare the size characteristics of our database and generated samples. 
The grain size distribution curve for the grains in the F50 database, computed using this method, is presented in Fig. ~\ref{fig:database_distributions}.

Other than qualitative visual inspection of the database and generated sand grains in this method, we also adopt the quantitative surface orientation metrics introduced by \citet{orosz2021surface}.
The surface orientation tensor is a computationally simple method proposed to characterize the form of irregular grains and grains used as load-bearing materials in structural and geotechnical engineering. 
It is based on the orientations of the faces of the grain using a special weighted tensor metric. 
This method provides measures of shape comparable to the oriented bounding box-based methods found in the literature \citep{o1985finding,gottschalk2000collision}.  The surface orientation tensor is essentially a fabric tensor \citep{satake1982fabric,satake1983fundamental} that belongs to the individual grain and is based on the outward normal vectors of the faces forming the grain surface and weighted proportionally to the faces' surface area. 
Its first eigenvector (which corresponds to the largest eigenvalue) can predict the grain's major contact orientation, and its eigenvalues can be used to characterize the grain form by computing measures of compactness $C$, flakiness $F$, and elongation $E$. 

The symmetric second order surface orientation tensor $f$ for a grain is defined as:
\begin{equation}
f_{i j} = \frac{1}{\sum_k A^k} \sum_k \left(A^k n_i^k n_j^k\right),
\label{eq:sot}
\end{equation}
where $A^k$ and $\tensor{n}$ are the area and the outward unit normal vector respectively of the $k$-th planar face of a polyhedral convex or concave grain surface.
The compactness $C$, flakiness $F$, and elongation $E$ metrics can be then calculated as:
\begin{equation}
C:=\frac{f_3}{f_1},
F:=\frac{f_1-f_2}{f_1}, \text{ and }
E:=\frac{f_2-f_3}{f_1},
\label{eq:CFE}
\end{equation}
where are $f_1,f_2,f_3$ are the eigenvalues of the surface orientation tensor in decreasing order.
$C,F,E$ add up to unity and can be presented in a ternary plot. 
These surface orientation metrics have been selected as a quantitative measure of the database and generated grains in this work due to the ease of implementation and presentation.
Other shape metrics that will consider more complex surface textures will be considered in future work.
The $C,F,E$ metrics were considered adequate to capture the current resolution of the F50 sand grain scans and provide a comprehensive matrix for comparison between grains.
The surface orientation metrics for the F50 sand database of this work is demonstrated in Fig.~\ref{fig:database_distributions}.


\subsection{Pre-processing of the point cloud data}
\label{sec:processing_point_cloud_data}

In order to prepare the point cloud dataset for training, we utilize the F50 sand grains dataset as described in the previous section. This dataset is comprised of 1,551 sand grains, providing a rich source of data for our model. Instead of directly using the surface meshes, we opt to represent each sand grain by the vertices of its corresponding surface mesh, thereby transforming the data into point clouds as described in Section~\ref{sec:autoencoder_theory}.

A majority of the surface meshes in the dataset have 600 vertices, with 1,542 sand grains featuring this exact number of points. However, there are 9 surface meshes that have fewer than 600 points. To maintain consistency and simplicity in our dataset, we decide to remove these sand grains with fewer points from the dataset. This decision is based on the assumption that these grains are outliers within the original dataset and thus do not contribute significantly to the overall pattern and structure that the autoencoder should learn. By streamlining the dataset in this manner, we ensure a more uniform and consistent input for training our point cloud autoencoder.

We pre-process the data set for training by processing each grain point cloud individually. 
Firstly, we transform the coordinates of every point cloud so that its center is located at the origin (0, 0, 0). 
This standardization enables our model to focus on the shapes of the sand grains without the influence of their original positions in space.
We also scale every grain point cloud coordinate system using the maximum dimension of all the grains in the data set so that all the point coordinates are in the (-1,1) range.
All the grains are scaled identically -- the sense of relative scale across the data set is maintained and the generated grains will be transformed back to the original scale.

To enrich the features stored in the point cloud data structure, we utilize the underlying unweighted undirected graph that represents each surface mesh. 
For every grain, we leverage this graph to generate node embeddings using the node2vec algorithm \citep{grover2016node2vec}.
 Node2vec is a graph embedding technique that learns continuous feature representations for nodes in networks by optimizing a feature learning task based on random walks. In our implementation, each graph node/ point is represented by an 8-feature embedding.
By combining these embeddings with the three coordinates of each point, we form an 11-dimensional input feature set for the point cloud autoencoder. 
The random walks used in the node2vec algorithm help encode information about the adjacency of points within each cloud, providing valuable context for the autoencoder. 
It is important to note that the decoder in our autoencoder architecture does not reconstruct these additional features; its primary purpose is to predict the spatial coordinates of the points. 
However, these enriched features are utilized to enhance the training of the encoder, as will be described in the following section.

As a final step in pre-processing the dataset, we perform data augmentation to increase the amount and diversity of the available data for training and validation. This is achieved by randomly rotating each grain in the dataset about the origin (0, 0, 0). We apply a rotation using three random Euler angles for each axis, with angle ranges between $-\pi/2$ and $\pi/2$ for each of the x, y, and z axes. The rotation matrix is then computed from these Euler angles.
After determining the centroid of the point cloud, we subtract it from the points to center the point cloud at the origin. Then, we apply the computed rotation matrix to the centered points. This rotation is performed once for every grain in the dataset, effectively doubling the data available for training and validating the autoencoder from 1,542 to 3,084 point clouds.
It is important to note that the node2vec embeddings are considered constant, as they are solely based on the node connectivity. The augmented dataset will be used exclusively for ensuring a more robust training of the autoencoder. In contrast, only the original 1,542 grains and their corresponding embeddings will be used in training the denoising diffusion model. This approach is taken to ensure that the denoising diffusion model learns not just the shapes of individual grains, but also the exact distribution of grains present in the original database.

\section{Point cloud denoising diffusion for sand grain generation}
\label{sec:denoising_diffusion_results}

In this section, we discuss the implementation and training of the neural network architectures of this work.
In Section~\ref{sec:autoencoder_training}, we provide the details for the point cloud convolutional autoencoder architecture, the training results, as well as the point cloud embedding representation and reconstruction.
In Section~\ref{sec:denoising_diffusion_training}, we present the training of the denoising diffusion algorithm, the results of the grain generation, and the comparison of the generated grains to those in the database.

\subsection{Training of point cloud autoencoder}
\label{sec:autoencoder_training}

In this section, we focus on training an autoencoder for point cloud data using a point cloud 1D convolution approach, similar to the method proposed in \citep{achlioptas2018learning} as described in Section~\ref{sec:autoencoder}. 
Although alternative autoencoder architectures could have been considered, we opted for the point cloud convolution due to its computational efficiency. Euclidean grid-based 3D autoencoders were not employed because of the resolution-related challenges associated with 3D voxels. 
On the other hand, graph convolutional autoencoders, which do not suffer from resolution issues, could have been utilized by converting the grain surface meshes into node-weighted undirected graphs. However, we decided against this approach as predicting the connectivity of the mesh graph would add unnecessary computational overhead for the point cloud reconstruction task. 
Nonetheless, we retain some of the graph information by enriching the point cloud input features with the node2vec algorithm, effectively incorporating valuable topological insights into the autoencoder training process.

The point cloud autoencoder is designed to process point cloud data samples with 11 input features, where 3 correspond to the coordinates and the remaining 8 are derived from a node2vec algorithm as a pre-processing step as described in Section~\ref{sec:processing_point_cloud_data}. The architecture comprises an encoder and a decoder, which together facilitate the embedding and reconstruction of the input point cloud data.

\begin{figure}[h!]
  \centering
  \begin{minipage}{0.30\textwidth}
    \centering
    \includegraphics[width=0.9\textwidth, angle=0]{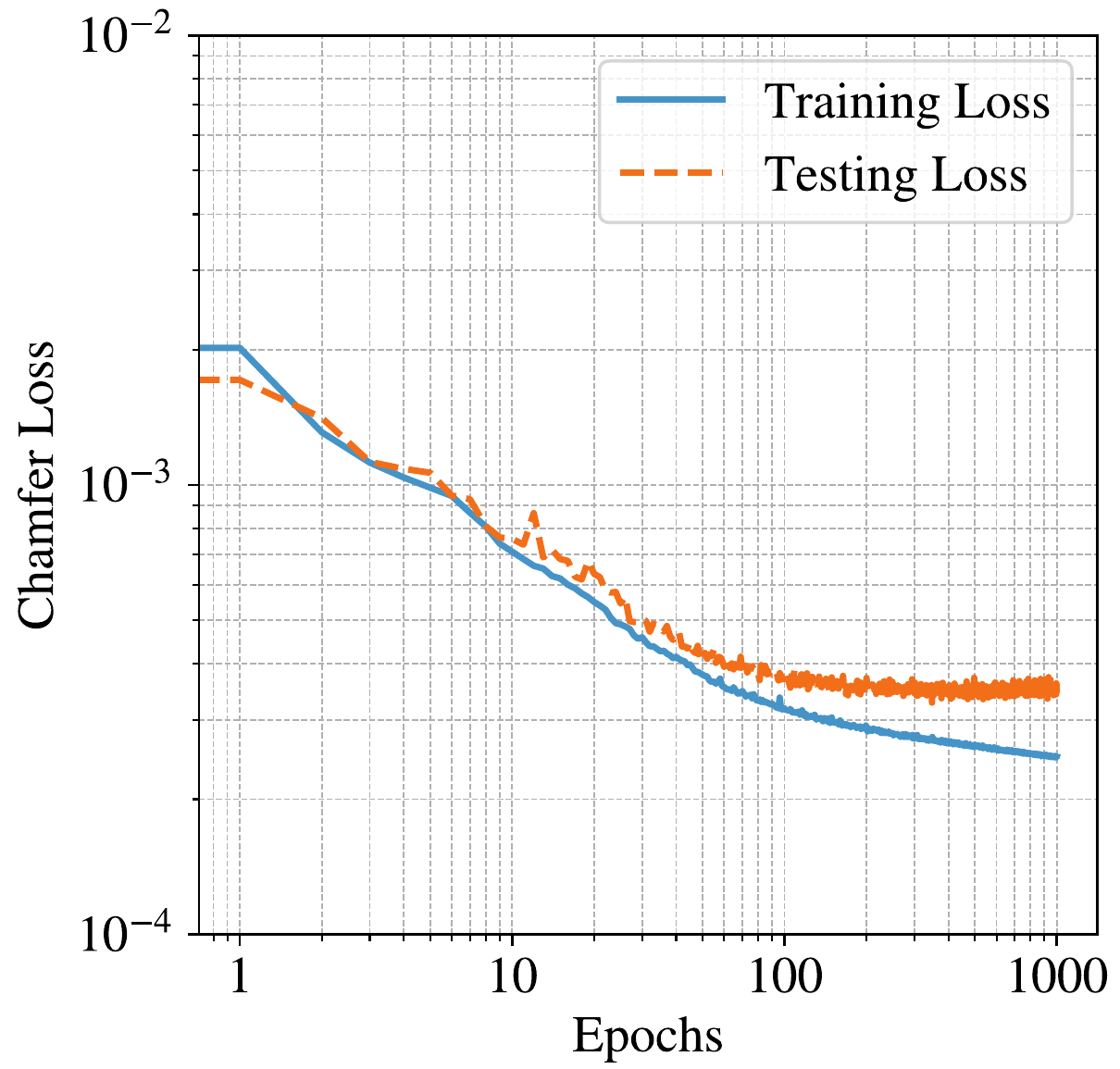}
  \end{minipage}
  \hfill 
  \begin{minipage}{0.6\textwidth}
    \centering
    \includegraphics[width=0.90\textwidth, angle=0]{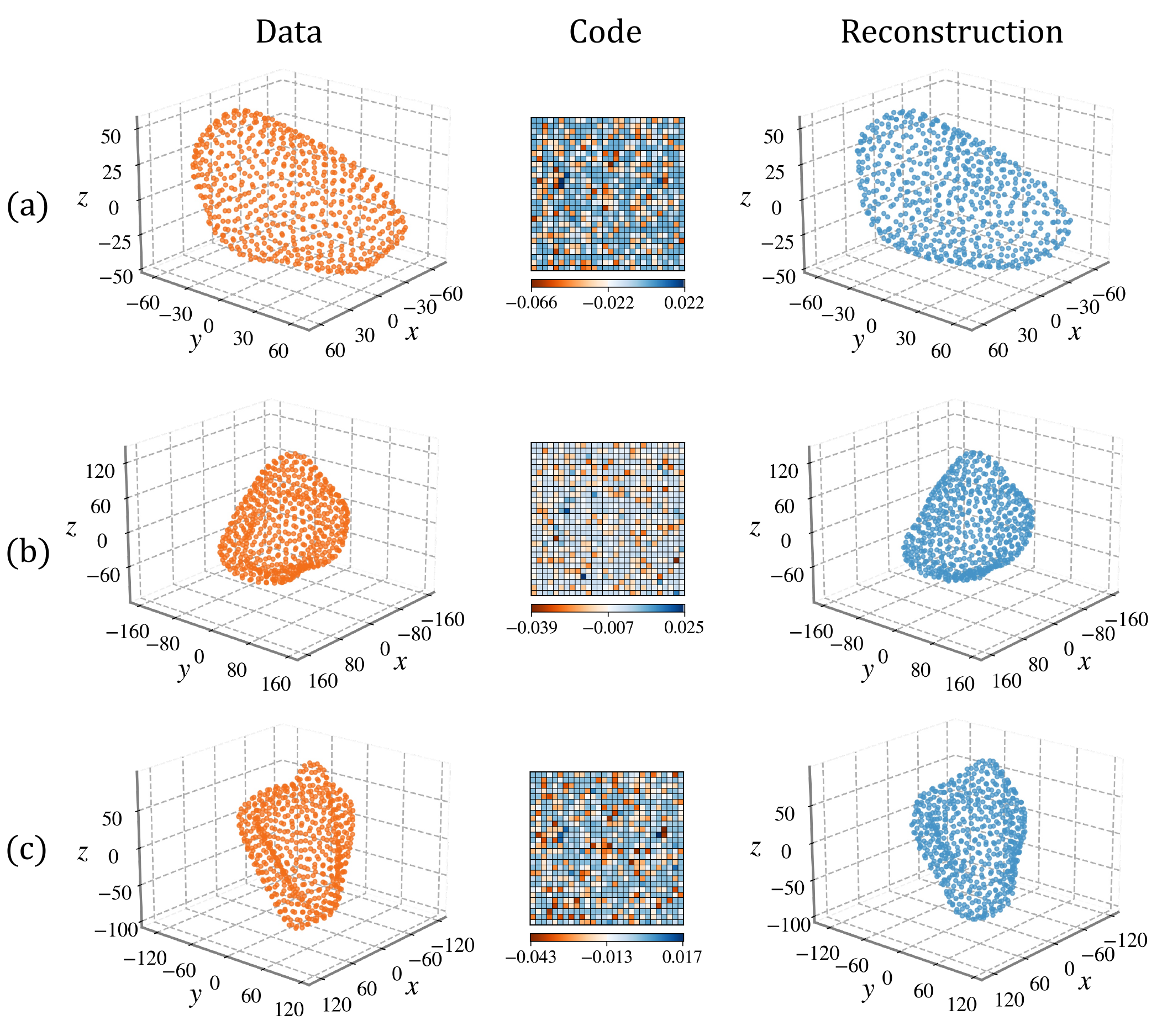}
  \end{minipage}
\caption{Left: Training and testing Chamfer loss function for the point cloud autoencoder.
Right: Point cloud embedding and reconstruction generated by the trained grain point cloud autoencoder for three data samples.}
\label{fig:autoencoder_results}
\end{figure}

Upon inputting a point cloud, the data first passes through the encoder. The encoder begins with an initial 1D convolutional layer that takes the 11 input features and transforms them into 64 channels. This transformation is followed by a Parametric Rectified Linear Unit (PReLU) activation function to introduce non-linearity.  Subsequently, the data go through a second 1D convolutional layer that expands the 64 channels into 784 channels, followed by another PReLU activation function. These point-wise convolutional layers in the encoder serve to extract and transform relevant features from the input point cloud data.
Once the data have passed through the pointwise convolutional layers, they are processed by an AdaptiveMaxPool1d layer, which compresses the data into a compact representation of dimension 28x28 (with one feature channel). This compact representation, or embedding, captures the underlying structure and geometry of the input point cloud while reducing its dimensionality. 
This embedded representation of the point cloud will be utilized in the denoising diffusion algorithm.

The embedded data is then passed to the decoder, which aims to reconstruct the original point cloud. The decoder starts with a 1D convolutional layer that takes the 784-dimensional embedding and reduces it to 64 channels. This is followed by a PReLU activation function to maintain non-linearity. 
Finally, the decoder employs another 1D convolutional layer that generates an output of 1800 channels, which is then reshaped into the original point cloud dimensions of (600,3).

To optimize the model, the Adam optimization algorithm is employed with an initial learning rate of 0.001. This learning rate is dynamically adjusted during training using a scheduler, specifically the ReduceLROnPlateau method. The scheduler monitors the model's performance and reduces the learning rate by a factor of 0.99 if no improvement is observed in the loss function for 5 consecutive epochs. This adaptive approach enables the model to find an optimal solution with a balance between exploration and exploitation during training.
The autoencoder is trained using the Chamfer Distance 3D loss function of Eq.~\eqref{eq:chamfer}.
In our implementation, all the point clouds have the same number of points ($|\tensor{P}_A| = |\tensor{P}_B|= 600$).

The training of the autoencoder is conducted on 2467 point cloud samples of the augmented database (Section~\ref{sec:processing_point_cloud_data}) and validating on 617 samples.
The training lasts for 1000 epochs with a batch size of 16 samples.
The results of the training are shown in Fig.~\ref{fig:autoencoder_results} including the comparison of the training and testing Chamfer loss curves. 
Three random samples of input point clouds from the testing set,  the $28\times 28$ embedding, and their reconstructions are also demonstrated.
The reconstructions are in good agreement in overall shape and size compared with the input grains.
There is a minor smoothening effect observed in some of the sharper features and edges of the reconstructed point clouds as expected.
These features could be better captured by further augmenting the dataset, increasing the size of the representation, or the size of the autoencoder architecture
but the reconstruction capacity was deemed satisfactory for the scope of this work; and more detail-oriented architectures will be considered in future work.

After the training and validation of the autoencoder,  we process every point cloud in the database through the encoder architecture to generate the point cloud embeddings $\tensor{x}$.
Thus, we generate 3084 point cloud embeddings of $28\times 28$ dimensions that will be used to train the denoising diffusion algorithm as described in the following section.


\subsection{Training of denoising diffusion model and generation results}
\label{sec:denoising_diffusion_training}

In this section, we discuss the training of the denoising diffusion algorithm described in Section~\ref{sec:denoising_architecture} for generating realistic sand grains. The algorithm is trained on the 1,542 point cloud embeddings derived from the encoder, as detailed in the previous section. This training process aims to enable the generation of a diverse set of sand grains, capturing the variations in shape and size observed in the original dataset.

In our work, we deploy a more compact version of the U-Net architecture used by \citet{nichol2021improved}, which suffices for our task of synthesizing the latent space of sand grains. 
Our U-Net comprises two stacks of layers that execute downsampling and upsampling over four stages each, with each stage hosting a residual block. 
The design incorporates one attention head with attention resolutions at 32$\times$32, 16$\times$16, and 8$\times$8. 
To generate the latent space embedding of the point clouds, which will subsequently be fed into the decoder architecture, we adapt the output of our network to produce a single output image channel. 
In terms of channel widths from higher to lower resolutions, our U-Net uses a scheme of $[C,2C,4C,8C]$, with the base model channel size chosen as $C=32$. 
The architecture's conditioning on the diffusion time step $t$ is achieved through an embedded feature vector. 
This feature vector is the output of a two-layer sequential network, featuring two dense layers of 128 neurons each, coupled with a Sigmoid Linear Unit (SiLU) and a Linear activation function, respectively.
To train the diffusion model, we perform 100,000 training steps on the 1,542 samples of point cloud embeddings. We adopt 1,000 diffusion time steps with a linear noise schedule, as used in \cite{ho2020denoising}. An Adam optimizer \citep{kingma2014adam} is employed, with a learning rate of $10^{-4}$ and a batch size of 128.

\begin{figure}[h!]
\centering
\includegraphics[width=.70\textwidth ,angle=0]{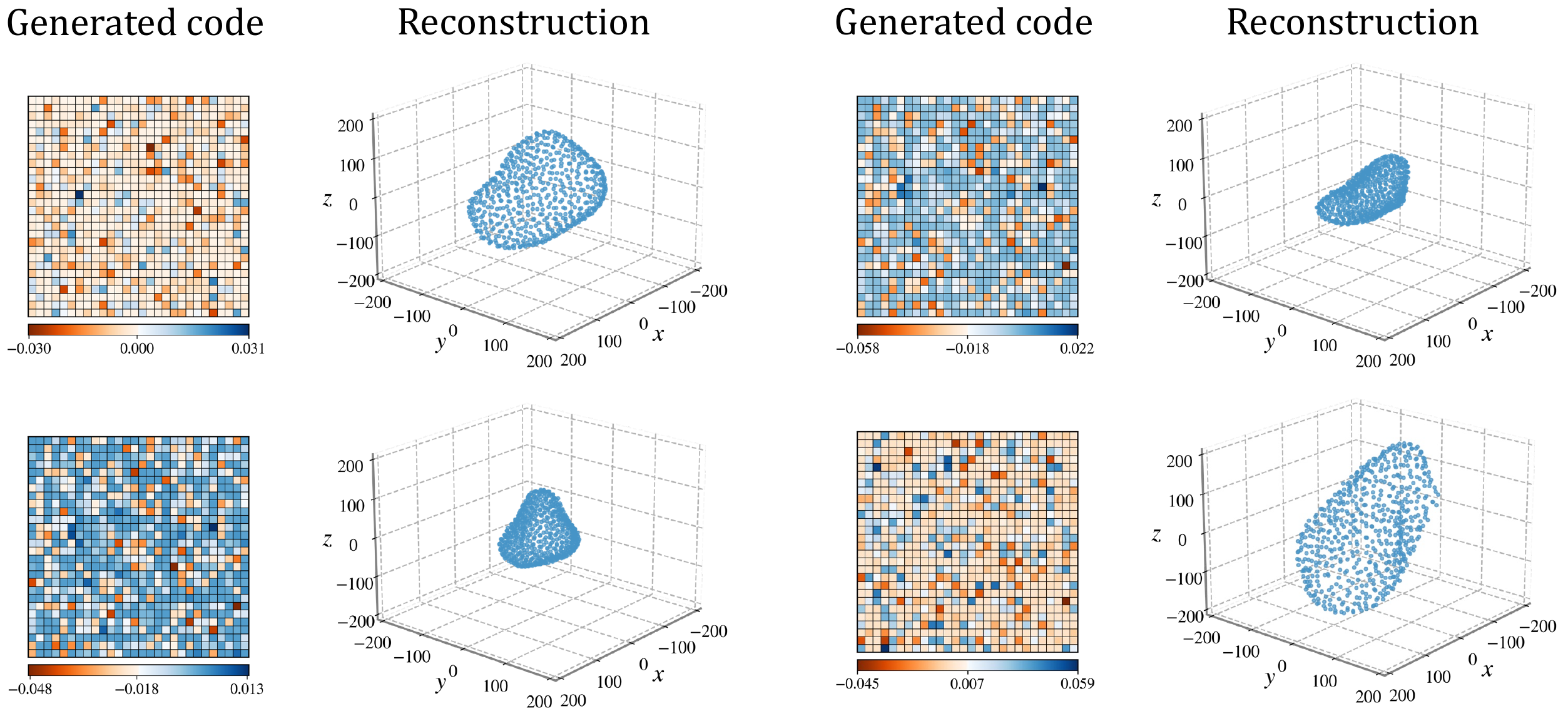} 
\caption{Generated point cloud embeddings from the trained diffusion algorithm and their corresponding reconstructions by the point cloud decoder.}
\label{fig:diffusion_results}
\end{figure}

\begin{figure}[h!]
\centering
\includegraphics[width=.95\textwidth ,angle=0]{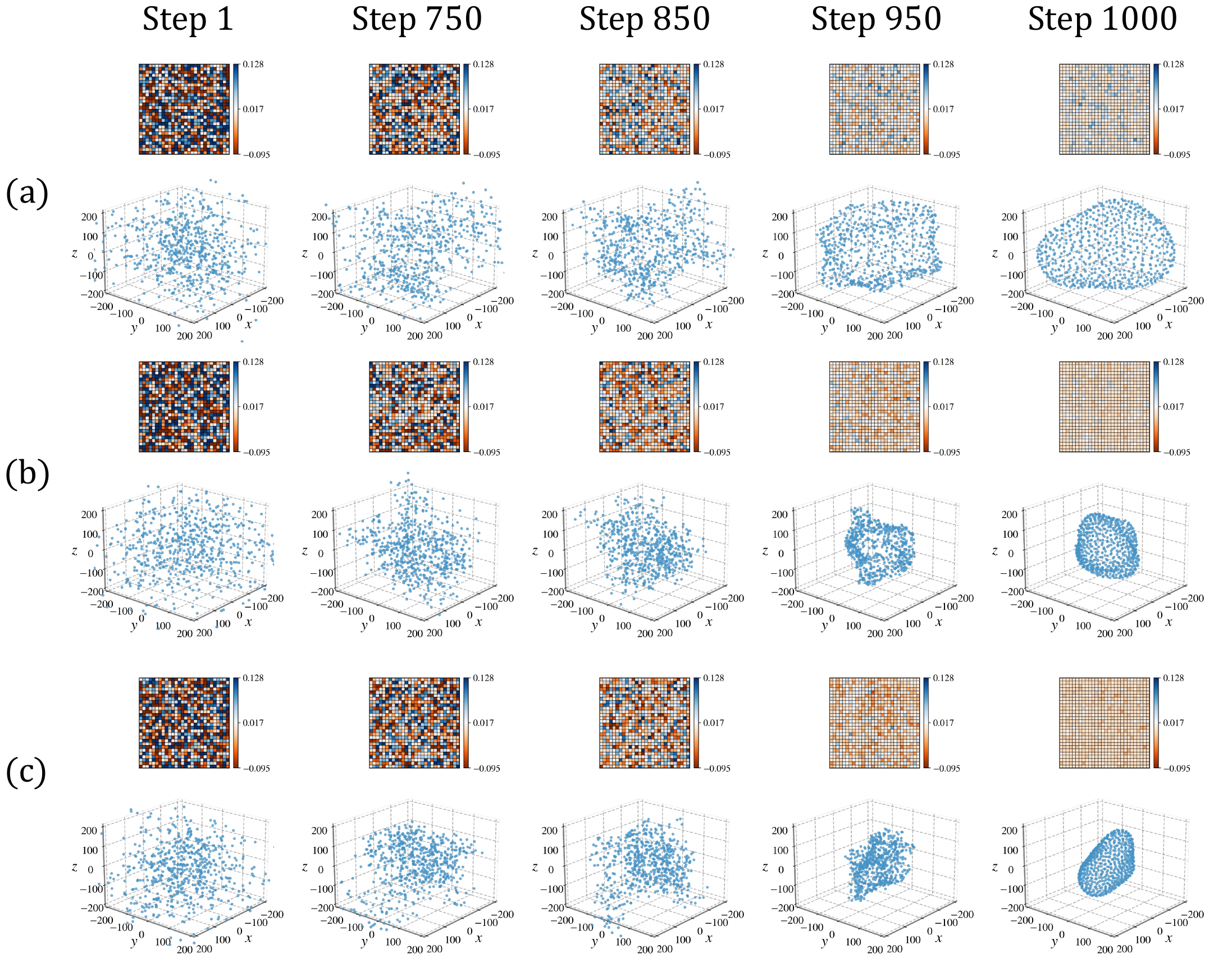} 
\caption{Generated point cloud embeddings and their corresponding reconstructions for the 0,  750, 850,  and 1000 denoising diffusion time steps.}
\label{fig:diffusion_steps}
\end{figure}

In Fig.~\ref{fig:diffusion_results}, we showcase the results obtained by sampling from the trained diffusion model. Four generated embeddings are illustrated, along with their corresponding point clouds, which are reconstructed by passing the embeddings through the decoder architecture. The reconstructed point clouds are plotted at the same scale, and we can observe that the algorithm is capable of generating sand grains with a variety of shapes and sizes.
The iterative process of the denoising algorithm is illustrated in Fig.~\ref{fig:diffusion_steps}, which presents the generated embeddings and their corresponding reconstructed point clouds at diffusion steps 1, 750, 850, 950, and 1,000. Convergence of the reconstructed point clouds is observed to commence around the 750th step in most cases, while the point clouds generated prior to this step appear predominantly as random noise.

To further assess the quality of the generated point clouds as sand grains, we construct surface meshes to visualize the reconstructed volumes. The PyMeshLab open-source library \citep{pymeshlab} is utilized for this purpose. We first estimate the point cloud normals using the 5 nearest neighbors for each point, followed by the application of the ball pivoting algorithm \citep{bernardini1999ball} to reconstruct the surface with a triangular mesh. To ensure complete reconstructed surfaces, we also apply a filter to close any discontinuities on the surface mesh.

\begin{figure}[h!]
\centering
\includegraphics[width=.80\textwidth ,angle=0]{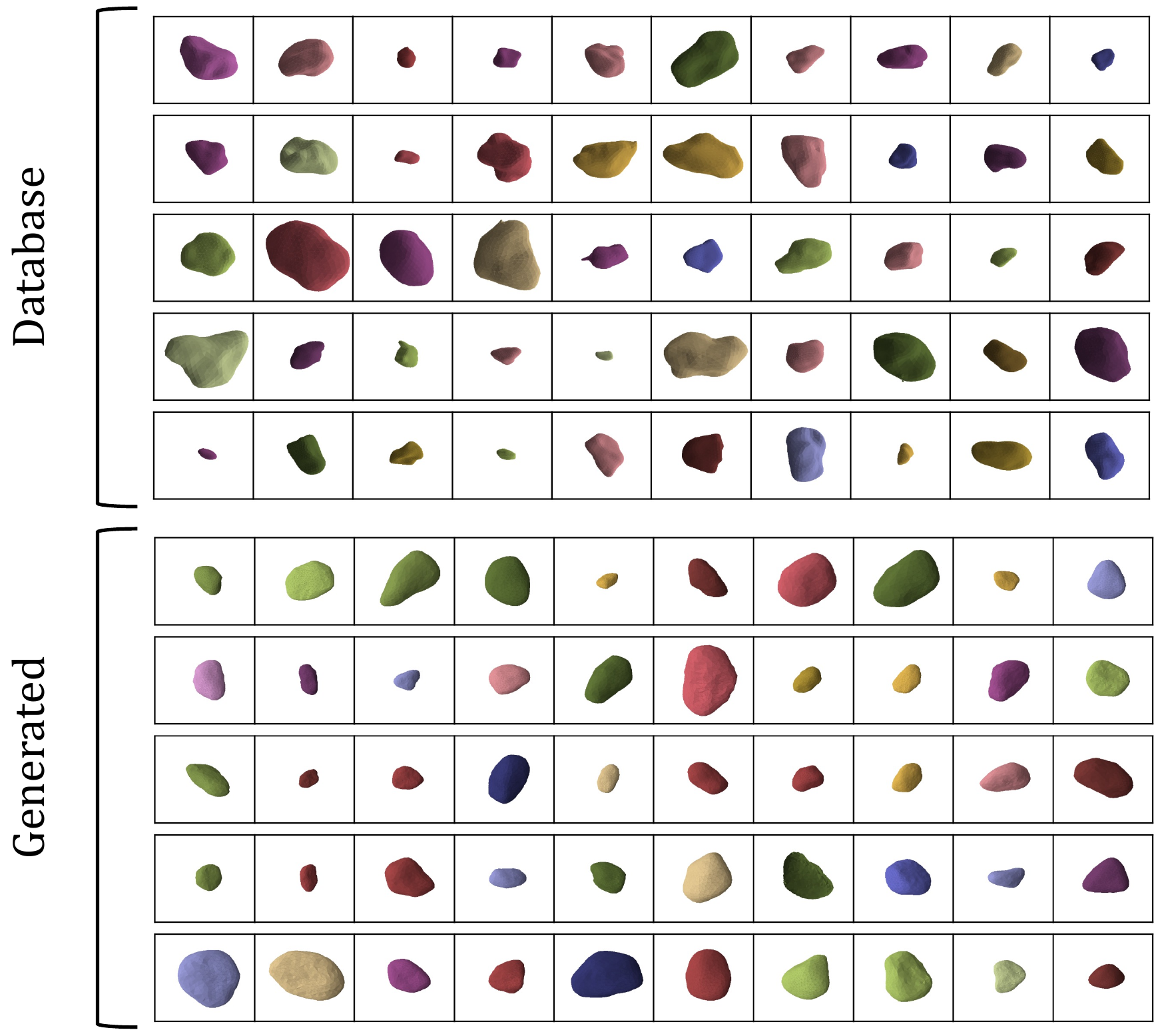} 
\caption{Comparison of 50 randomly sampled database and 50 generated sand grains.}
\label{fig:grain_comparison}
\end{figure}

The surface reconstruction results are presented in Fig.~\ref{fig:grain_comparison}, where 50 randomly sampled database sand grains are compared with 50 generated and meshed grains produced by the diffusion algorithm. A visual inspection reveals that the generated grains are in good agreement with those present in the original database, indicating the efficacy of the algorithm in capturing the essential characteristics of sand grains. In the following section, we present a more comprehensive analysis of the size and shape distributions of the sand grains, further demonstrating the capabilities of the denoising diffusion algorithm in generating realistic sand grain point clouds with properties consistent to those in the original data set.


\section{Generation of granular assemblies}
\label{sec:granular_assemblies}

In this  section, we explore the capabilities of the denoising diffusion algorithm in generating not only individual sand grains but also entire sand grain assemblies. 
This is an essential step towards simulating realistic granular materials and understanding their collective behavior. 
Section~\ref{sec:size_shape} presents sampling experiments of the diffusion algorithm, comparing the size and shape property distributions of the generated assemblies to those found in the original database, thereby assessing the algorithm's ability to produce diverse and representative assemblies. 
In Section~\ref{sec:gravity_deposition} we demonstrate how the generated grains can be sub-sampled and filtered to construct assemblies with targeted property distributions, showcasing the flexibility and potential applications of our approach in various granular material studies and engineering problems.

\subsection{Size and shape analysis}
\label{sec:size_shape}

In this section, we conduct sampling experiments using the diffusion algorithm to investigate whether the generated samples accurately capture the size and shape property distributions of the original database as a whole. The aim of these experiments is to assess the diffusion algorithm's ability to produce representative sand grains and assemblies for granular material studies.

We perform two sampling experiments, generating two different sample sizes. In the first experiment, we generate 1,536 samples, a number closely matching the 1,542 original sand grains in the database. In the second experiment, we generate a significantly larger sample size of 50,000 grains to evaluate the algorithm's ability to maintain the original sample properties when generating an extensive collection of sand grains. We choose to generate the grains batch-wise, with a batch size of 128, to efficiently utilize computational resources.

The sampling of one batch takes approximately 36 seconds to complete. Consequently, the first sampling experiment, with 1,536 samples, takes around 7 minutes to finish, while the second experiment, involving 50,000 grains, takes approximately 235 minutes. It is worth noting that the generation process can be parallelized to accelerate the generation time, making it feasible for larger-scale applications and studies involving massive granular assemblies.

\begin{figure}[h!]
\centering
\includegraphics[width=.70\textwidth ,angle=0]{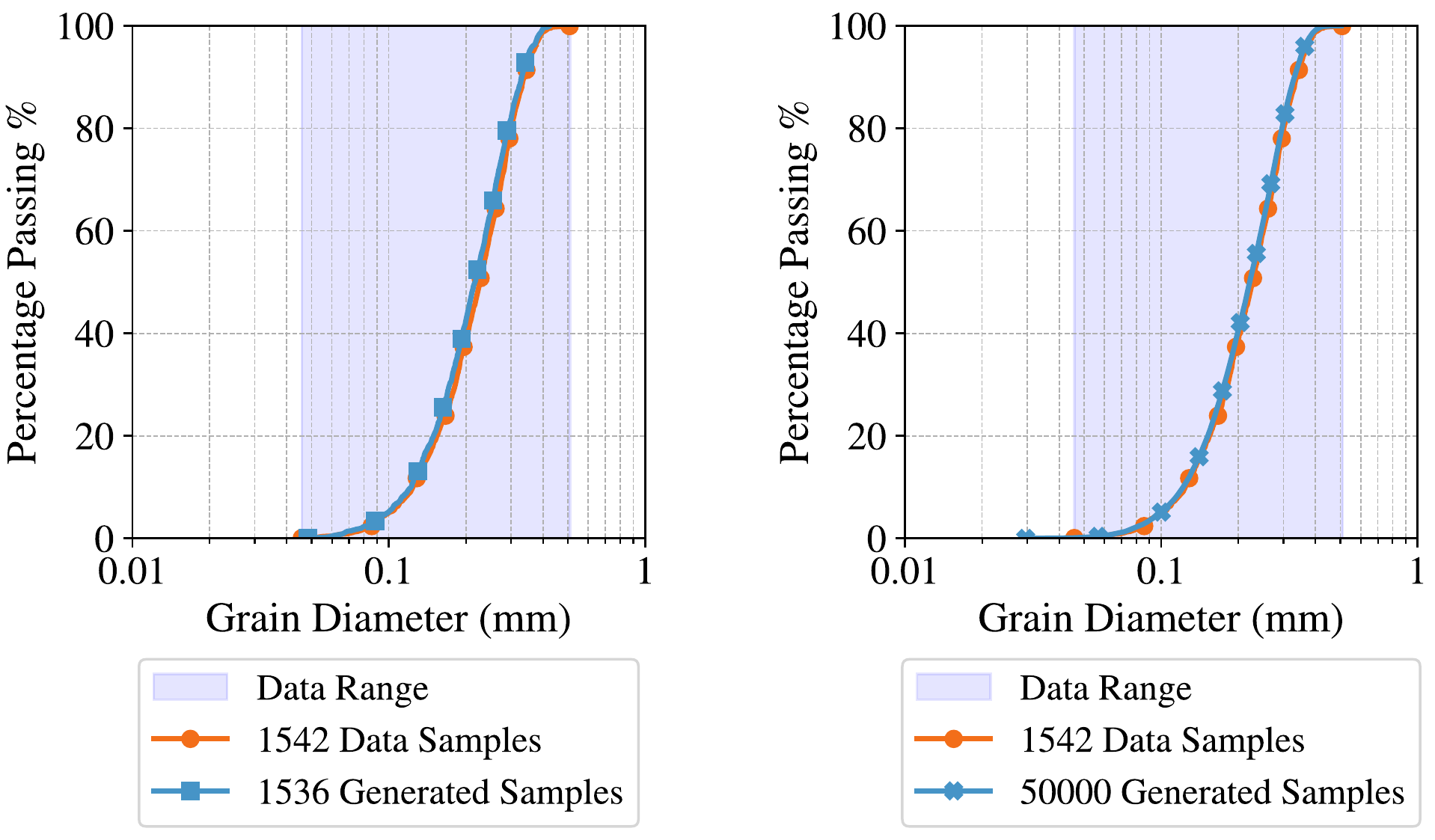} 
\caption{Grain size distribution comparison of the database and generated grains.  
Left: Distribution for 1,536 generated samples -- none of which were out of the database grain size range.
Right: Distribution of 50,000 grains -- 58 ($0.116\%$) of which are out of the database grain size range. }
\label{fig:size_distributions}
\end{figure}

Upon generating the point clouds, we proceed to postprocess each sample individually to gather the size and shape metrics for the entire set of samples. To obtain diameter information for each grain, we fit every point cloud to a minimum sphere. Additionally, we mesh every point cloud following the methodology described in the previous section and calculate the surface orientation tensor to determine the shape metrics C, E, and F for each grain, as detailed in Section~\ref{sec:grain_metrics}.

\begin{table}[htbp]
\centering
\begin{tabular}{|c|c|c|c|c|c|}
\hline
Sample & $D_{10}$ (mm) & $D_{30}$ (mm) & $D_{60}$ (mm) & $C_u$ & $C_c$ \\
\hline
Database & 0.1241 & 0.1819 & 0.2503 & 2.0166 & 1.0652 \\
1,536 Samples & 0.1222 & 0.1725 & 0.2404 & 1.9677 & 1.0136 \\
50,000 Samples & 0.1218 & 0.1762 & 0.2455 & 2.0163 & 1.0385 \\
\hline
\end{tabular}

\vspace{0.5cm}

\begin{tabular}{|c|c|c|c|}
\hline
Sample & Mean (mm) & Median (mm) & Standard Deviation (mm) \\
\hline
Database & 0.2292 & 0.2269 & 0.0808 \\
1,536 Samples & 0.2212 & 0.2174 & 0.0785 \\
50,000 Samples & 0.2242 & 0.2214 & 0.0787 \\
\hline
\end{tabular}

\vspace{0.5cm}

\begin{tabular}{|c|c|c|c|c|}
\hline
\parbox{1.8cm}{\centering Sample} & Min Diameter (mm) & Max Diameter (mm) & \parbox{2.5cm}{\centering Grains out of\ the data range} & \parbox{2.5cm}{\centering Percentage of grains out of the data range} \\
\hline
Database & 0.0458 & 0.5105 & -- & -- \\
1,536 Samples & 0.0483 & 0.4255 & 0 & 0.0\% \\
50,000 Samples & 0.0296 & 0.5035 & 58 & 0.116\% \\
\hline
\end{tabular}
\caption{Grain size metrics comparison of the database and generated grain samples.}
\label{tab:grain_size_metrics}
\end{table}

The size metrics for the two grain samples, as well as those for the original database, are presented in the grain size distribution curves depicted in Fig. ~\ref{fig:size_distributions}. 
The curves for the two generated distributions appear to be in good agreement with the data ranges.
The results of the grain size metrics comparison between the database and the generated grain samples are also presented in Table~\ref{tab:grain_size_metrics}. The table is divided into three parts, each providing different aspects of the size metrics.
The first part of the table reports the $D_{10}$, $D_{30}$, and $D_{60}$ values, which represent the diameters corresponding to 10\%, 30\%, and 60\% of the cumulative grain size distribution, respectively, along with the coefficients of uniformity ($C_u$) and curvature ($C_c$). The results indicate that both the 1,536 samples and the 50,000 samples are able to closely match the database values.
The second part of the table presents the mean, median, and standard deviation of the grain diameters for each sample set. 
The third part of the table shows the minimum and maximum diameters for each sample set, as well as the number and percentage of grains that fall outside the original database's size range. It is worth noting that for the 1,536 samples, none of the generated grains fall outside the database's size range. For the larger 50,000-sample set, only 58 grains (0.116\%) are found to be outside the range, indicating that the diffusion algorithm remains consistent in generating realistic sand grains even when producing a significantly larger number of samples.

\begin{figure}[h!]
\centering
\includegraphics[width=.80\textwidth ,angle=0]{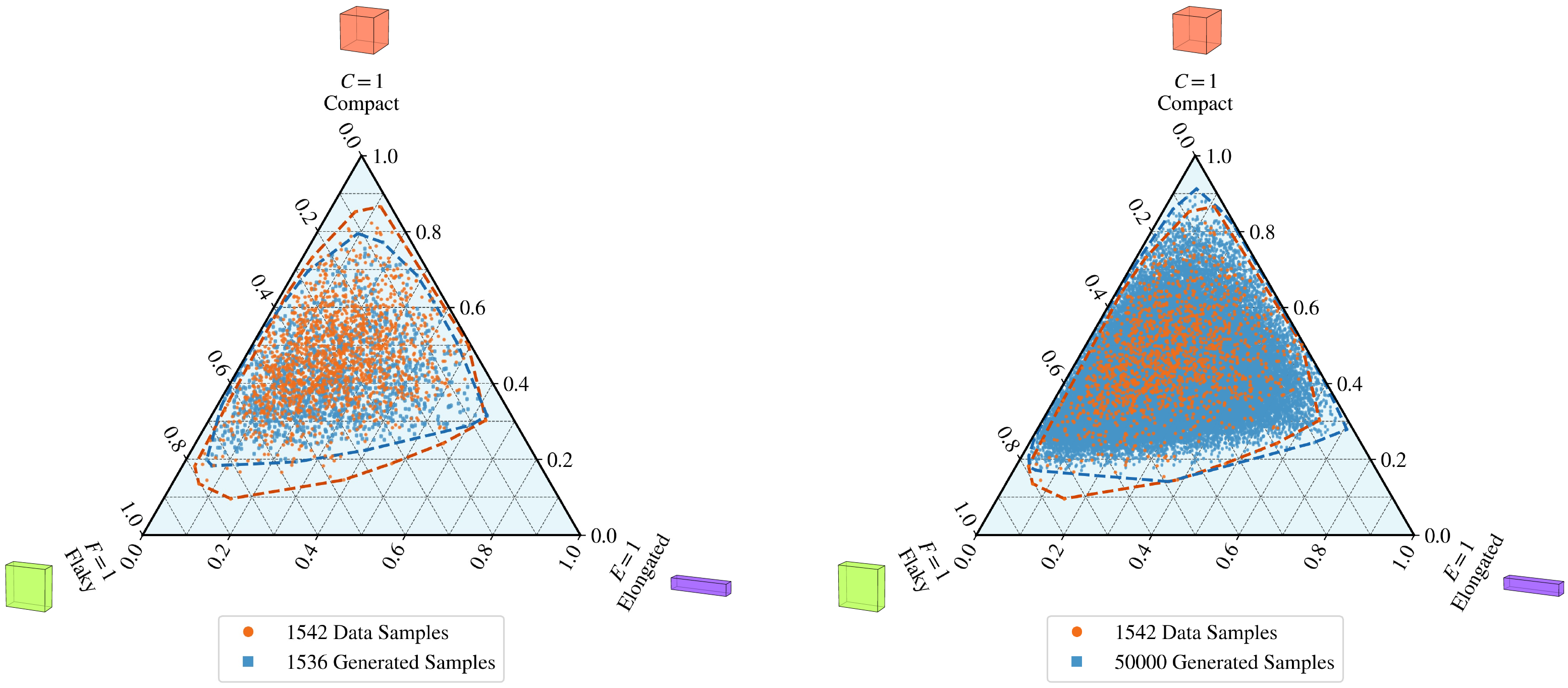} 
\caption{Grain shape distributions of database and generated grains for the 1,536 and 50,000 sampling experiments.
The convex hull of the distribution points are also demonstrated.}
\label{fig:shape_distributions}
\end{figure}

The results of the grain size metrics comparison between the database and the generated grain samples, as presented in Table~\ref{tab:grain_size_metrics}, provide a closer insights into the performance of the diffusion algorithm. When sampling a number of grains (1,536) close to the number of grains in the original database (1,542), the algorithm demonstrates good agreement in the size distribution metrics, indicating that we can effectively generate a size distribution twin of the original dataset.
Furthermore, when generating a significantly larger sample of 50,000 grains, the results show that the size distribution metrics statistically converge to those of the original database. This demonstrates that the diffusion algorithm is capable of generating samples larger than the original dataset without compromising the size distribution characteristics found in the original sand grains. 

\begin{table}[htbp]
\centering

\begin{tabular}{|c|c|c|c||c|c|c||c|c|c|}
\hline
Sample & $C_\text{mean}$ & $F_\text{mean}$ & $E_\text{mean}$& $C_\text{median}$ & $F_\text{median}$ & $E_\text{median}$& $C_\text{st. dev.}$ & $F_\text{st. dev.}$ & $E_\text{st. dev.}$  \\
\hline
Database & 0.464 & 0.328 & 0.208 & 0.456 & 0.325 & 0.195 & 0.119 & 0.145 & 0.108  \\
1,536 Samples & 0.415 & 0.360 & 0.225 & 0.400 & 0.362 & 0.212 & 0.110 & 0.150 & 0.121  \\
50,000 Samples & 0.417 & 0.357 & 0.226 & 0.399 & 0.358 & 0.211 & 0.399 & 0.358 & 0.211  \\
\hline
\end{tabular}

\caption{Grain shape metrics comparison of the database and generated grain samples.}
\label{tab:grain_shape_metrics}
\end{table}

The grain shape metrics comparison between the database and the generated grain samples is presented in Figure~\ref{fig:shape_distributions} and Table~\ref{tab:grain_shape_metrics}. The figure shows the distributions of the compactness ($C$), flakiness ($F$), and elongation ($E$) coefficients as described in Section~\ref{sec:sand_grain_dataset}. The convex hull of the distribution points for all three shape metrics demonstrates a good match between the original database and the generated samples. However, it is worth noting that a few shape outliers present in the original database were not fully captured by the diffusion algorithm. 

The grain shape metrics in Table~\ref{tab:grain_shape_metrics} are also in good agreement between the original database and the generated samples, both for the 1,536 and 50,000 grain sampling experiments. These results indicate that the diffusion algorithm is capable of generating samples with a similar range of shape properties as the original database. 
Similar to the size distribution results, the shape metrics demonstrate that the diffusion algorithm is capable of generating larger samples without losing the original shape distribution characteristics. This highlights the robustness of the algorithm and its potential in generating sand grain assemblies with targeted size and shape properties that will be discussed in following section.


\subsection{Targeted sand grain assembly generation}
\label{sec:gravity_deposition}

In this section, we demonstrate the potential of our diffusion algorithm in generating sand grain assemblies with targeted properties by visualizing gravity deposition simulations. One of the key advantages of our method is the ability to generate grains not just individually but also as part of larger assemblies. This capability enables us to quickly generate assemblies that either mirror the properties of the original database or consist of sub-sampled distributions of the original data properties.

\begin{figure}[h!]
\centering
\includegraphics[width=.99\textwidth ,angle=0]{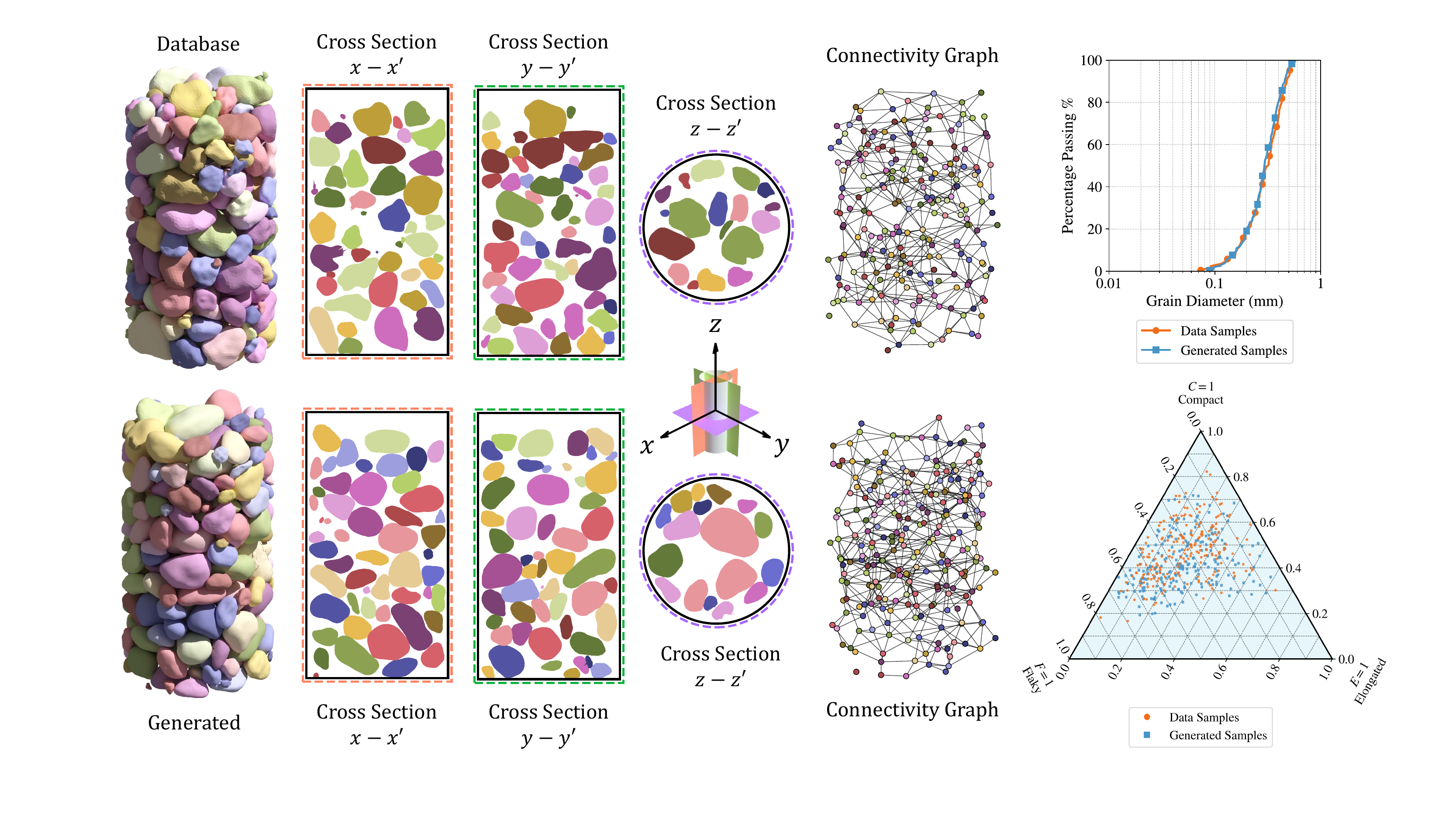} 
\caption{Comparison of database and generated randomly sampled granular assemblies and cross-sections along $x,y$ and $z$ axes.
The connectivity graphs, grain size and shape distributions are also demonstrated.}
\label{fig:cross_sections}
\end{figure}

The fast and mass generation of grains allows us to create assemblies with specific characteristics. By generating assemblies with targeted properties, it is possible to study the behavior of granular materials in various conditions, including those that are not present in the original database. 
This is particularly useful for understanding the influence of grain size and shape on the macroscopic behavior of granular materials and for simulating real-world scenarios that involve complex interactions between grains. The control and correlation of grain size and shape distributions to macroscale mechanical properties will be considered in future work.

Through the visualization of gravity deposition simulations, we demonstrate the effectiveness of our approach in generating assemblies with desired size and shape distributions.  The simulations are performed using the Blender physics engine \citep{blender}. 
It is important to note that our primary goal in this work is not to accurately reproduce the entire physics behavior of the grains but rather to provide a rendering of the assemblies in an equilibrium state once the deposition process is complete. To achieve this, the physics engine is employed for the gravity deposition and surface contact detection between grains.

The grains are given surface meshes, as described in the previous section, and a cylindrical container is defined for collision detection and to confine the generated grains. The container has a height of 2.5 mm and a diameter of 1.4 mm. 
To set up the simulations, grains are initially placed in random positions over the container. Specifically, the grains are positioned at random heights within the 7-14 mm range and randomly inside the 0-1.4 mm diameter range. The grains are then allowed to drop from their initial heights.
The simulation concludes when the grains reach the 2.5 mm mark at the top of the cylindrical container. The number of grains necessary to fill the container up to this height is not predetermined, as we may sample different size distributions. 

\begin{table}[htbp]
\centering

\begin{tabular}{|c|c|c|c|c|c|}
\hline
Sample & Porosity & CN & Transitivity & Graph density & Local efficiency \\
\hline
Database & 0.402 & 4.804 & 0.223 & 0.023 & 0.346 \\
\hline
Generated & 0.364 & 5.494 & 0.245 & 0.023 & 0.436 \\
\hline
\end{tabular}

\caption{Porosity and graph properties comparison of database and generated assemblies of Fig.~\ref{fig:cross_sections}.}
\label{tab:sample_graph_properties}
\end{table}

In our first assembly generation experiment, we aim to compare the database grains with those generated using the denoising diffusion algorithm. 
To achieve this, we randomly sampled grains from the database until the 2.5 mm mark in the cylindrical container is reached. Likewise, we continuously generate grains using the denoising diffusion algorithm until the container is filled up to the same height. 
It is important to note that no specific filters or constraints have been applied to the generated grain assemblies at this stage.
In this experiment, it took 209 grains from the database to fill the container, while 237 generated grains were required to achieve the same fill level. 
The results of this comparison are illustrated in Fig.~\ref{fig:cross_sections}, where we present 3D renders of both assemblies. Furthermore, we provide cross-sectional views of the assemblies, taken diametrically along the $x$ and $y$ axes and passing through the middle point of the cylinder for the $z$ axis.  We also display the grain size and shape distributions for the two samples, which show good agreement, even for a small sub-sample of the distributions. 

For a closer analysis of the assemblies, we construct the connectivity graph for both the database and generated assemblies, with the results displayed in Fig.~\ref{fig:cross_sections}. Within these graphs, each node represents a sand grain and an edge is shared between nodes (grains) that are in contact. The graph is undirected -- there is no significance in the direction of the edge. 
Moreover, we present a range of metrics derived from these connectivity graphs in Table~\ref{tab:sample_graph_properties}. 
The porosity is slightly higher in the database assembly (0.402) compared to the generated one (0.364). 
The coordination number (CN), a measure of the average number of contacts between grains, is marginally higher in the generated assembly (5.494) than in the database assembly (4.804). 
The transitivity, graph density, and local efficiency, which provide insights into the network connectivity within the assemblies, also demonstrate relatively close values between the database and generated assemblies.
It's important to note, however, that the differences between the properties of the two assemblies are not significant and can be considered satisfactory for the scope of the current grain deposition simulations. 
To gain a clearer understanding of the grain and assembly contact metrics, larger sample sizes and multiple simulation repetitions would be beneficial in providing average metrics. This level of in-depth study, while not included in this current work, will form part of our future research. 
We plan to delve deeper into the physics underlying the generated grains and assemblies, to further enhance the accuracy and effectiveness of our approach.

\begin{figure}[h!]
\centering
\includegraphics[width=.9\textwidth ,angle=0]{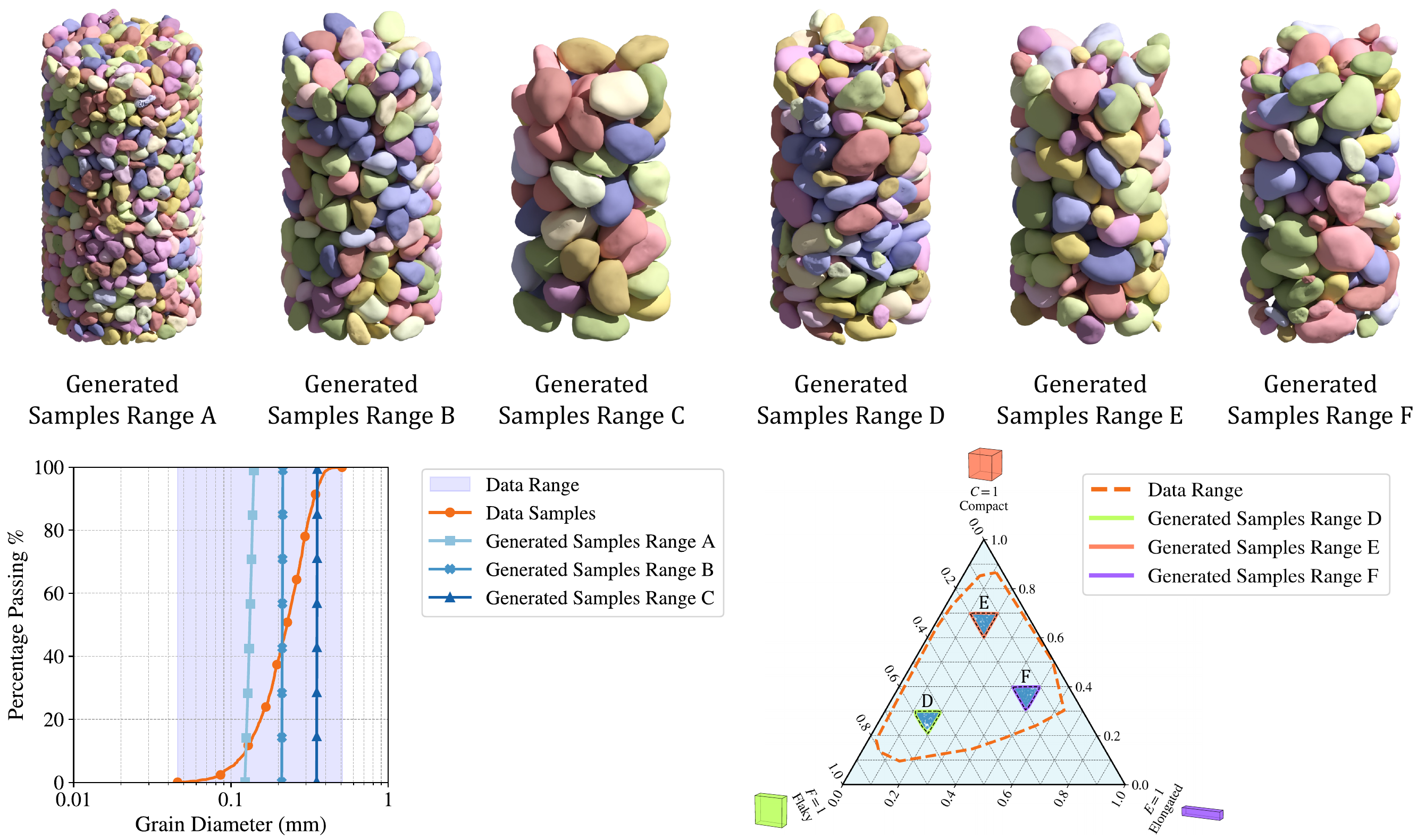} 
\caption{LEFT: Selective generation of three uniform grain size distribution ranges of increasing grain sizes.
Range A contains grains of 0.1225-0.1400 mm diameter, range B contains 0.2100-0.2135 mm diameter, and range C contains 0.3500-0.3535 mm diameter.
The original database grain size distribution is also shown for comparison.
RIGHT: Selective generation of three uniform grain shape distribution ranges D,E,  and F for different surface orientation tensor metrics.
The original database grain shape distribution boundaries are also shown for comparison.}
\label{fig:six_ranges}
\end{figure}

In the following assembly generation experiments, we focus on generating assemblies with targeted size and shape distributions. The results of these experiments are presented in Fig.~\ref{fig:six_ranges}. We begin by generating three assemblies with specific targeted size ranges, selecting three uniform size ranges to achieve this. These selective generation experiments involve creating assemblies with three uniform grain size distribution ranges, each containing grains of increasing sizes.
We filter out grains from the 50,000 generated grain sampled database and we perform the gravity deposition simulations until the 2.5 mm mark is reached.
Range A includes grains with diameters between 0.1225 and 0.1400 mm, range B comprises grains with diameters between 0.2100 and 0.2135 mm, and range C features grains with diameters between 0.3500 and 0.3535 mm. 
It required 1,580 grains for range A, 331 for range B, and 64 for range C to fill the container.
For comparison purposes, the original database grain size distribution is also displayed alongside these targeted ranges. It is important to note that the chosen ranges are subsampled from the original database range. As demonstrated in the previous section, very few outliers outside these ranges can be generated, and even when they are generated, there is limited confidence that they will exhibit realistic shape properties. 

\begin{figure}[h!]
\centering
\includegraphics[width=.9\textwidth ,angle=0]{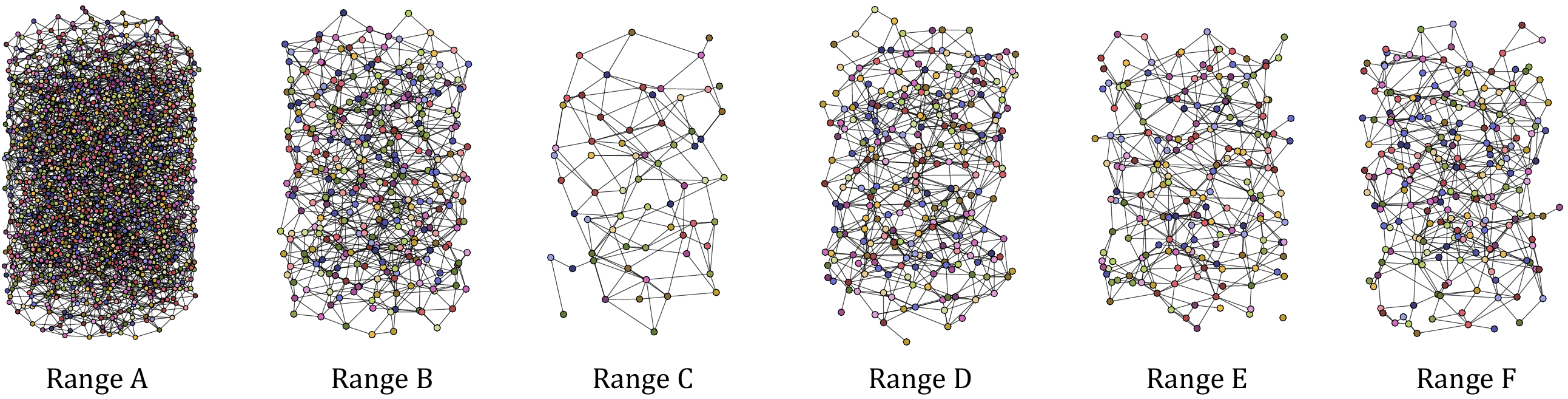} 
\caption{Grain connectivity graphs of generated assemblies of Fig.~\ref{fig:six_ranges} corresponding to target size ranges A,B,C and target shape ranges D,E, and F.}
\label{fig:six_ranges_graphs}
\end{figure}

\begin{table}[htbp]
\centering

\begin{tabular}{|c|c|c|c|c|c|}
\hline
Sample & Porosity & CN & Transitivity & Graph density & Local efficiency \\
\hline
Range A & 0.305 & 8.381 & 0.318 & 0.005 & 0.579 \\
\hline
Range B & 0.366 & 6.671 & 0.279 & 0.020 & 0.460 \\
\hline
Range C & 0.451 & 4.281 & 0.243 & 0.068 & 0.288 \\
\hline
\hline
Range D & 0.385 & 5.812 & 0.276 & 0.023 & 0.472 \\
\hline
Range E & 0.389 & 4.869 & 0.235 & 0.026 & 0.413 \\
\hline
Range F & 0.388 & 5.283 & 0.229 & 0.025 & 0.396 \\
\hline
\end{tabular}

\caption{Porosity and graph properties comparison of Ranges A-F of Fig.~\ref{fig:six_ranges}.}
\label{tab:sample_graph_properties_six}
\end{table}

Lastly, we generate assemblies with targeted uniform shape range properties. To achieve this, we subsample grains from the 50,000 generated samples in three triangular ranges of the surface orientation tensor plot, ensuring that they fall within the boundaries of the database shape distribution metrics C, F, and E.
Range D features grains with compactness metric C within [0.2, 0.3], flakiness metric F within [0.5, 0.6], and elongation metric E within [0.1, 0.2]. Range E contains grains with metrics C within [0.6, 0.7], range F within [0.1, 0.2], and range E within [0.1, 0.2]. Range F consists of grains with metrics with C within [0.3, 0.4],  F within [0.1, 0.2], and E within [0.4, 0.5].
To fill the container up to the 2.5 mm mark, 255 grains were required for range D, 191 grains for range E, and 212 grains for range F.

The connectivity graph and assembly metric results of these simulations are also presented in Fig.~\ref{fig:six_ranges_graphs} and Table~\ref{tab:sample_graph_properties_six}.
For Ranges A-C, which were targeted based on grain size, we observed an inverse relationship between the grain size and the number of grains, or equivalently, the number of nodes in the corresponding graph. As an example, Range A, with the smallest grain sizes, populated the cylindrical container with a densely packed network of 1,580 nodes, while in contrast, Range C, with the larger grain sizes, accommodated only 64 nodes in a more sparse configuration.
This variation in grain sizes is reflected in the assembly metrics. For instance, the porosity, representing the void space in the assembly, increases from Range A (0.305) to Range C (0.451), indicating that assemblies with larger grain sizes tend to have more void space. Similarly, the coordination number, a measure of the average number of contacts per grain, decreases with increasing grain size, from 8.381 in Range A to 4.281 in Range C. This trend suggests that larger grains have fewer points of contact, likely due to their reduced quantity in the assembly and looser packing.
For Ranges D-F, we targeted different shape characteristics. Interestingly, despite the variations in grain shape, the volumetric properties of these ranges led to comparable assembly metrics. Each of these ranges required a similar number of grains to fill the container, reflecting in their almost identical porosities and other graph properties. 
However, the subtle differences in assembly metrics across Ranges D-F hint at the potential role of grain shape in determining the contact mechanics and macroscopic behavior of assemblies. A more detailed analysis is required to fully understand this role, which is a focus of our future work with a more robust grain contact physics simulator.

The targeted generation of grain assemblies with specific size and shape properties offers a wide range of potential applications across various engineering fields. 
Although the current work does not directly address the mechanical properties of sand grains and the assemblies they form, this will be an essential part of future research. For example, future studies could focus on the effect of packing and target packing of voids in assemblies to better understand the mechanical behavior of granular materials under different stress conditions.
Generating subsampled assembly twins can provide valuable insights into material properties, especially when the original scanned database does not contain sufficient data points. Running simulations with these generated twins can yield useful information that is otherwise unattainable from the original database. 
For instance, the original database had only 90 grains in the 0.1225-0.1400 mm diameter range -- 5.83\% of the original database.  
However, we were able to generate 1,580 grains in this size range (Range A), significantly expanding the possibilities for studying the macroscale properties of finer grains.
Note that the 50,000 grain sample has 2784 in the 0.1225-0.1400 mm diameter range -- 5.57\% of the generated data.
With the ability to generate larger grain assembly twins that mimic the size and shape distribution properties of the original database, researchers can delve deeper into the behavior and characteristics of the studied granular databases.


\section{Conclusion}
\label{sec:conclucion}

We introduce a novel method for generating sand grain assemblies with targeted size and shape properties using denoising diffusion probabilistic models. 
We effectively embed the point cloud representations of the F50 sand grains to create a lower dimensional space with a convolutional autoencoder.
We then perform the denoising diffusion generation in the latent space to generate embedding that can then be decoded and preprocessed into sand grains.
Our approach effectively captures the characteristics present in the original database -- not only for every generated grain individually but also preserving the size and shape distributions of the training data.
We showcase the method's potential by generating grain assemblies through targeted sampling and gravity deposition simulations, providing insights into possible applications in geotechnical engineering, concentrated solar power, explosive materials,  and beyond. Future work will consider additional properties, such as mechanical behaviors and packing characteristics, to further understand granular material behavior at the macroscale.
The generation of grain assemblies using quick and versatile diffusion algorithms facilitates the development of digital twins and synthetic databases in engineering applications where data availability may be limited.

\section{Acknowledgments}
The authors are primarily supported by the Department of Energy, National Nuclear Security Administration, Predictive Science Academic Alliance Program (PSAAP) under Award Number DE-NA0003962 
Additional support for NN Vlassis and WC Sun are provided by 
 the National Science Foundation under grant contracts CMMI-1846875 and OAC-1940203, and
 the Dynamic Materials and Interactions Program from the Air Force Office of Scientific 
Research under grant contracts FA9550-19-1-0318,  FA9550-21-1-0391 and FA9550-21-1-0027.
The experiments that obtained the microCT images utilized resources of the Advanced Photon Source (APS); a U.S. Department of Energy
(DOE) Office of Science User Facility operated for the DOE Office of Science by Argonne
National Laboratory (ANL) under Contract No. DE-AC02-06CH11357. The SMT images
presented in this paper were collected using the x-ray Operations and Research Beamline Station
13-BMD ANL. The authors would like to thank Dr. Mark Rivers of APS for his help in
performing the SMT scans. The authors also acknowledge the support of GeoSoilEnviroCARS
(Sector 13), which is supported by the National Science Foundation, Earth Sciences (EAR-
1128799), and the U.S. Department of Energy (DOE), Geosciences Contracts No DE-FG02-
94ER14466 and DE-AC02-06CH11357.
These supports are gratefully acknowledged. 
The views and conclusions contained in this document are those of the authors, 
and should not be interpreted as representing the official policies, either expressed or implied, 
of the sponsors, including the Army Research Laboratory or the U.S. Government. 
The U.S. Government is authorized to reproduce and distribute reprints for 
Government purposes notwithstanding any copyright notation herein.

\section{Data and code availability}
The computer code that support the findings of this study are available from the corresponding author upon request. 

\section{CRediT authorship contribution statement}
Nikolaos Vlassis: Conceptualization, Methodology, Software, Validation, 
Formal analysis, Investigation, Data Curation, Writing – Original Draft. WaiChing Sun: Validation, Resource, Writing – Original Draft, Supervision, Project administration, Funding acquisition. Khalid A. Alshibli: Data Curation, Resources, Formal analysis, Writing  – Review \& Editing; Richard Regueiro, Writing  – Review \& Editing, Supervision, Project ddministration, Funding acquisition. 
 
\section{Compliance with ethical standards}
The authors declare that they have no conflict of interest.

\bibliographystyle{plainnat}
\bibliography{main}

\begin{thebibliography}{79}
\providecommand{\natexlab}[1]{#1}
\providecommand{\url}[1]{\texttt{#1}}
\expandafter\ifx\csname urlstyle\endcsname\relax
  \providecommand{\doi}[1]{doi: #1}\else
  \providecommand{\doi}{doi: \begingroup \urlstyle{rm}\Url}\fi

\bibitem[Achlioptas et~al.(2018)Achlioptas, Diamanti, Mitliagkas, and
  Guibas]{achlioptas2018learning}
Panos Achlioptas, Olga Diamanti, Ioannis Mitliagkas, and Leonidas Guibas.
\newblock Learning representations and generative models for 3d point clouds.
\newblock In \emph{International conference on machine learning}, pages 40--49.
  PMLR, 2018.

\bibitem[Ahmed and Martinez(2021)]{ahmed2021triaxial}
Sheikh~Sharif Ahmed and Alejandro Martinez.
\newblock Triaxial compression behavior of 3d printed and natural sands.
\newblock \emph{Granular Matter}, 23:\penalty0 1--21, 2021.

\bibitem[Altuhafi et~al.(2016)Altuhafi, Coop, and
  Georgiannou]{altuhafi2016effect}
Fatin~N Altuhafi, Matthew~R Coop, and Vasiliki~N Georgiannou.
\newblock Effect of particle shape on the mechanical behavior of natural sands.
\newblock \emph{Journal of Geotechnical and Geoenvironmental Engineering},
  142\penalty0 (12):\penalty0 04016071, 2016.

\bibitem[Beiser(2019)]{beiser2019world}
Vince Beiser.
\newblock Why the world is running out of sand.
\newblock \emph{BBC Future}, 18, 2019.

\bibitem[Bernardini et~al.(1999)Bernardini, Mittleman, Rushmeier, Silva, and
  Taubin]{bernardini1999ball}
Fausto Bernardini, Joshua Mittleman, Holly Rushmeier, Cl{\'a}udio Silva, and
  Gabriel Taubin.
\newblock The ball-pivoting algorithm for surface reconstruction.
\newblock \emph{IEEE transactions on visualization and computer graphics},
  5\penalty0 (4):\penalty0 349--359, 1999.

\bibitem[Bowman et~al.(2001)Bowman, Soga, and Drummond]{bowman2001particle}
Elisabeth~T Bowman, Kenichi Soga, and W~Drummond.
\newblock Particle shape characterisation using fourier descriptor analysis.
\newblock \emph{Geotechnique}, 51\penalty0 (6):\penalty0 545--554, 2001.

\bibitem[Bronstein et~al.(2017)Bronstein, Bruna, LeCun, Szlam, and
  Vandergheynst]{bronstein2017geometric}
Michael~M Bronstein, Joan Bruna, Yann LeCun, Arthur Szlam, and Pierre
  Vandergheynst.
\newblock Geometric deep learning: going beyond euclidean data.
\newblock \emph{IEEE Signal Processing Magazine}, 34\penalty0 (4):\penalty0
  18--42, 2017.

\bibitem[Buarque~de Macedo et~al.(2018)Buarque~de Macedo, Marshall, and
  Andrade]{buarque2018granular}
Robert Buarque~de Macedo, Jason~P Marshall, and Jos{\'e}~E Andrade.
\newblock Granular object morphological generation with genetic algorithms for
  discrete element simulations.
\newblock \emph{Granular Matter}, 20:\penalty0 1--12, 2018.

\bibitem[Chen et~al.(2020)Chen, Zhang, Zen, Weiss, Norouzi, and
  Chan]{chen2020wavegrad}
Nanxin Chen, Yu~Zhang, Heiga Zen, Ron~J Weiss, Mohammad Norouzi, and William
  Chan.
\newblock Wavegrad: Estimating gradients for waveform generation.
\newblock \emph{arXiv preprint arXiv:2009.00713}, 2020.

\bibitem[Cil and Alshibli(2015)]{cil2015modeling}
Mehmet~B Cil and Khalid~A Alshibli.
\newblock Modeling the influence of particle morphology on the fracture
  behavior of silica sand using a 3d discrete element method.
\newblock \emph{Comptes Rendus M{\'e}canique}, 343\penalty0 (2):\penalty0
  133--142, 2015.

\bibitem[Community(2018)]{blender}
Blender~Online Community.
\newblock Blender - a 3d modelling and rendering package, 2018.
\newblock URL \url{http://www.blender.org}.

\bibitem[Dafalias and Manzari(2004)]{dafalias2004simple}
Yannis~F Dafalias and Majid~T Manzari.
\newblock Simple plasticity sand model accounting for fabric change effects.
\newblock \emph{Journal of Engineering mechanics}, 130\penalty0 (6):\penalty0
  622--634, 2004.

\bibitem[{Dawson-Haggerty et al.}()]{trimesh}
{Dawson-Haggerty et al.}
\newblock trimesh.
\newblock URL \url{https://trimsh.org/}.

\bibitem[Druckrey et~al.(2016)Druckrey, Alshibli, and
  Al-Raoush]{druckrey20163d}
Andrew~M Druckrey, Khalid~A Alshibli, and Riyadh~I Al-Raoush.
\newblock 3d characterization of sand particle-to-particle contact and
  morphology.
\newblock \emph{Computers and Geotechnics}, 74:\penalty0 26--35, 2016.

\bibitem[Duriez and Bonelli(2021)]{duriez2021precision}
J{\'e}r{\^o}me Duriez and St{\'e}phane Bonelli.
\newblock Precision and computational costs of level set-discrete element
  method (ls-dem) with respect to dem.
\newblock \emph{Computers and Geotechnics}, 134:\penalty0 104033, 2021.

\bibitem[Elbaz et~al.(2017)Elbaz, Avraham, and Fischer]{elbaz20173d}
Gil Elbaz, Tamar Avraham, and Anath Fischer.
\newblock 3d point cloud registration for localization using a deep neural
  network auto-encoder.
\newblock In \emph{Proceedings of the IEEE conference on computer vision and
  pattern recognition}, pages 4631--4640, 2017.

\bibitem[Gadelha et~al.(2018)Gadelha, Wang, and
  Maji]{gadelha2018multiresolution}
Matheus Gadelha, Rui Wang, and Subhransu Maji.
\newblock Multiresolution tree networks for 3d point cloud processing.
\newblock In \emph{Proceedings of the European Conference on Computer Vision
  (ECCV)}, pages 103--118, 2018.

\bibitem[Gottschalk(2000)]{gottschalk2000collision}
Stefan~Aric Gottschalk.
\newblock \emph{Collision queries using oriented bounding boxes}.
\newblock The University of North Carolina at Chapel Hill, 2000.

\bibitem[Griffiths and Boehm(2019)]{griffiths2019review}
David Griffiths and Jan Boehm.
\newblock A review on deep learning techniques for 3d sensed data
  classification.
\newblock \emph{Remote Sensing}, 11\penalty0 (12):\penalty0 1499, 2019.

\bibitem[Grover and Leskovec(2016)]{grover2016node2vec}
Aditya Grover and Jure Leskovec.
\newblock node2vec: Scalable feature learning for networks.
\newblock In \emph{Proceedings of the 22nd ACM SIGKDD international conference
  on Knowledge discovery and data mining}, pages 855--864, 2016.

\bibitem[Guo et~al.(2020)Guo, Wang, Hu, Liu, Liu, and Bennamoun]{guo2020deep}
Yulan Guo, Hanyun Wang, Qingyong Hu, Hao Liu, Li~Liu, and Mohammed Bennamoun.
\newblock Deep learning for 3d point clouds: A survey.
\newblock \emph{IEEE transactions on pattern analysis and machine
  intelligence}, 43\penalty0 (12):\penalty0 4338--4364, 2020.

\bibitem[Gupta et~al.(2019)Gupta, Salager, Wang, and Sun]{gupta2019open}
Ritesh Gupta, Simon Salager, Kun Wang, and WaiChing Sun.
\newblock Open-source support toward validating and falsifying discrete
  mechanics models using synthetic granular materials---part i: Experimental
  tests with particles manufactured by a 3d printer.
\newblock \emph{Acta Geotechnica}, 14:\penalty0 923--937, 2019.

\bibitem[Hall et~al.(2010)Hall, Bornert, Desrues, Pannier, Lenoir, Viggiani,
  and B{\'e}suelle]{hall2010discrete}
Stephen~A Hall, Michel Bornert, Jacques Desrues, Yannick Pannier, Nicolas
  Lenoir, Gioacchino Viggiani, and Pierre B{\'e}suelle.
\newblock Discrete and continuum analysis of localised deformation in sand
  using x-ray $\mu$ct and volumetric digital image correlation.
\newblock \emph{G{\'e}otechnique}, 60\penalty0 (5):\penalty0 315--322, 2010.

\bibitem[Henkes and Wessels(2022)]{henkes2022three}
Alexander Henkes and Henning Wessels.
\newblock Three-dimensional microstructure generation using generative
  adversarial neural networks in the context of continuum micromechanics.
\newblock \emph{Computer Methods in Applied Mechanics and Engineering},
  400:\penalty0 115497, 2022.

\bibitem[Ho et~al.(2020)Ho, Jain, and Abbeel]{ho2020denoising}
Jonathan Ho, Ajay Jain, and Pieter Abbeel.
\newblock Denoising diffusion probabilistic models.
\newblock \emph{Advances in Neural Information Processing Systems},
  33:\penalty0 6840--6851, 2020.

\bibitem[Hooda et~al.(2022)Hooda, Pan, and Syed]{hooda2022survey}
Reetu Hooda, W~David Pan, and Tamseel~M Syed.
\newblock A survey on 3d point cloud compression using machine learning
  approaches.
\newblock \emph{SoutheastCon 2022}, pages 522--529, 2022.

\bibitem[Hyv{\"a}rinen and Dayan(2005)]{hyvarinen2005estimation}
Aapo Hyv{\"a}rinen and Peter Dayan.
\newblock Estimation of non-normalized statistical models by score matching.
\newblock \emph{Journal of Machine Learning Research}, 6\penalty0 (4), 2005.

\bibitem[Jefferies(1993)]{jefferies1993nor}
MG~Jefferies.
\newblock Nor-sand: a simle critical state model for sand.
\newblock \emph{G{\'e}otechnique}, 43\penalty0 (1):\penalty0 91--103, 1993.

\bibitem[Jerves et~al.(2017)Jerves, Kawamoto, and Andrade]{jerves2017geometry}
Alex~X Jerves, Reid~Y Kawamoto, and Jos{\'e}~E Andrade.
\newblock A geometry-based algorithm for cloning real grains.
\newblock \emph{Granular Matter}, 19:\penalty0 1--10, 2017.

\bibitem[Kamrin(2019)]{kamrin2019non}
Ken Kamrin.
\newblock Non-locality in granular flow: Phenomenology and modeling approaches.
\newblock \emph{Frontiers in Physics}, 7:\penalty0 116, 2019.

\bibitem[Kawamoto et~al.(2018)Kawamoto, And{\`o}, Viggiani, and
  Andrade]{kawamoto2018all}
Reid Kawamoto, Edward And{\`o}, Gioacchino Viggiani, and Jos{\'e}~E Andrade.
\newblock All you need is shape: Predicting shear banding in sand with ls-dem.
\newblock \emph{Journal of the Mechanics and Physics of Solids}, 111:\penalty0
  375--392, 2018.

\bibitem[Kingma and Ba(2014)]{kingma2014adam}
Diederik~P Kingma and Jimmy Ba.
\newblock Adam: A method for stochastic optimization.
\newblock \emph{arXiv preprint arXiv:1412.6980}, 2014.

\bibitem[Kingma and Welling(2013)]{kingma2013auto}
Diederik~P Kingma and Max Welling.
\newblock Auto-encoding variational bayes.
\newblock \emph{arXiv preprint arXiv:1312.6114}, 2013.

\bibitem[Kipf and Welling(2016)]{kipf2016variational}
Thomas~N Kipf and Max Welling.
\newblock Variational graph auto-encoders.
\newblock \emph{arXiv preprint arXiv:1611.07308}, 2016.

\bibitem[Kong et~al.(2020)Kong, Ping, Huang, Zhao, and
  Catanzaro]{kong2020diffwave}
Zhifeng Kong, Wei Ping, Jiaji Huang, Kexin Zhao, and Bryan Catanzaro.
\newblock Diffwave: A versatile diffusion model for audio synthesis.
\newblock \emph{arXiv preprint arXiv:2009.09761}, 2020.

\bibitem[Kuhn and Bagi(2004)]{kuhn2004contact}
Matthew~R Kuhn and Katalin Bagi.
\newblock Contact rolling and deformation in granular media.
\newblock \emph{International journal of solids and structures}, 41\penalty0
  (21):\penalty0 5793--5820, 2004.

\bibitem[Li et~al.(2018)Li, Zhang, Zhao, Burkhart, Brinson, and
  Chen]{li2018transfer}
Xiaolin Li, Yichi Zhang, He~Zhao, Craig Burkhart, L~Catherine Brinson, and Wei
  Chen.
\newblock A transfer learning approach for microstructure reconstruction and
  structure-property predictions.
\newblock \emph{Scientific reports}, 8\penalty0 (1):\penalty0 13461, 2018.

\bibitem[Lin and Ng(1997)]{lin1997three}
Xiaoshan Lin and T-T Ng.
\newblock A three-dimensional discrete element model using arrays of
  ellipsoids.
\newblock \emph{Geotechnique}, 47\penalty0 (2):\penalty0 319--329, 1997.

\bibitem[Liu et~al.(2019{\natexlab{a}})Liu, Sun, Li, Hu, and Wang]{liu2019deep}
Weiping Liu, Jia Sun, Wanyi Li, Ting Hu, and Peng Wang.
\newblock Deep learning on point clouds and its application: A survey.
\newblock \emph{Sensors}, 19\penalty0 (19):\penalty0 4188, 2019{\natexlab{a}}.

\bibitem[Liu et~al.(2019{\natexlab{b}})Liu, Han, Wen, Liu, and
  Zwicker]{liu2019l2g}
Xinhai Liu, Zhizhong Han, Xin Wen, Yu-Shen Liu, and Matthias Zwicker.
\newblock L2g auto-encoder: Understanding point clouds by local-to-global
  reconstruction with hierarchical self-attention.
\newblock In \emph{Proceedings of the 27th ACM International Conference on
  Multimedia}, pages 989--997, 2019{\natexlab{b}}.

\bibitem[Mandikal et~al.(2018)Mandikal, Navaneet, Agarwal, and
  Babu]{mandikal20183d}
Priyanka Mandikal, KL~Navaneet, Mayank Agarwal, and R~Venkatesh Babu.
\newblock 3d-lmnet: Latent embedding matching for accurate and diverse 3d point
  cloud reconstruction from a single image.
\newblock \emph{arXiv preprint arXiv:1807.07796}, 2018.

\bibitem[Menon and Elkan(2011)]{menon2011link}
Aditya~Krishna Menon and Charles Elkan.
\newblock Link prediction via matrix factorization.
\newblock In \emph{Machine Learning and Knowledge Discovery in Databases:
  European Conference, ECML PKDD 2011, Athens, Greece, September 5-9, 2011,
  Proceedings, Part II 22}, pages 437--452. Springer, 2011.

\bibitem[Mitchell et~al.(2005)Mitchell, Soga, et~al.]{mitchell2005fundamentals}
James~Kenneth Mitchell, Kenichi Soga, et~al.
\newblock \emph{Fundamentals of soil behavior}, volume~3.
\newblock John Wiley \& Sons New York, 2005.

\bibitem[Mollon and Zhao(2013)]{mollon2013generating}
Guilhem Mollon and Jidong Zhao.
\newblock Generating realistic 3d sand particles using fourier descriptors.
\newblock \emph{Granular Matter}, 15:\penalty0 95--108, 2013.

\bibitem[Moriguchi et~al.(2009)Moriguchi, Borja, Yashima, and
  Sawada]{moriguchi2009estimating}
Shuji Moriguchi, Ronaldo~I Borja, Atsushi Yashima, and Kazuhide Sawada.
\newblock Estimating the impact force generated by granular flow on a rigid
  obstruction.
\newblock \emph{Acta Geotechnica}, 4:\penalty0 57--71, 2009.

\bibitem[Muntoni and Cignoni(2021)]{pymeshlab}
Alessandro Muntoni and Paolo Cignoni.
\newblock {PyMeshLab}, January 2021.

\bibitem[Nichol and Dhariwal(2021)]{nichol2021improved}
Alexander~Quinn Nichol and Prafulla Dhariwal.
\newblock Improved denoising diffusion probabilistic models.
\newblock In \emph{International Conference on Machine Learning}, pages
  8162--8171. PMLR, 2021.

\bibitem[Nixon and Aguado(2019)]{nixon2019feature}
Mark Nixon and Alberto Aguado.
\newblock \emph{Feature extraction and image processing for computer vision}.
\newblock Academic press, 2019.

\bibitem[Nova and Wood(1979)]{nova1979constitutive}
R~Nova and David~Muir Wood.
\newblock A constitutive model for sand in triaxial compression.
\newblock \emph{International journal for numerical and analytical methods in
  geomechanics}, 3\penalty0 (3):\penalty0 255--278, 1979.

\bibitem[Orosz et~al.(2021)Orosz, Angelidakis, and Bagi]{orosz2021surface}
{\'A}kos Orosz, Vasileios Angelidakis, and Katalin Bagi.
\newblock Surface orientation tensor to predict preferred contact orientation
  and characterise the form of individual particles.
\newblock \emph{Powder Technology}, 394:\penalty0 312--325, 2021.

\bibitem[O'Rourke(1985)]{o1985finding}
Joseph O'Rourke.
\newblock Finding minimal enclosing boxes.
\newblock \emph{International journal of computer \& information sciences},
  14:\penalty0 183--199, 1985.

\bibitem[O'Sullivan and Bray(2017)]{o2017relating}
Catherine O'Sullivan and Jonathan~D Bray.
\newblock Relating the response of idealized analogue particles and real sands.
\newblock In \emph{numerical modeling in micromechanics via particle methods},
  pages 157--164. Routledge, 2017.

\bibitem[Peters et~al.(2005)Peters, Muthuswamy, Wibowo, and
  Tordesillas]{peters2005characterization}
JF~Peters, M~Muthuswamy, J~Wibowo, and A~Tordesillas.
\newblock Characterization of force chains in granular material.
\newblock \emph{Physical review E}, 72\penalty0 (4):\penalty0 041307, 2005.

\bibitem[Ramesh et~al.(2022)Ramesh, Dhariwal, Nichol, Chu, and
  Chen]{ramesh2022hierarchical}
Aditya Ramesh, Prafulla Dhariwal, Alex Nichol, Casey Chu, and Mark Chen.
\newblock Hierarchical text-conditional image generation with clip latents.
\newblock \emph{arXiv preprint arXiv:2204.06125}, 2022.

\bibitem[Rasul et~al.(2021)Rasul, Seward, Schuster, and
  Vollgraf]{rasul2021autoregressive}
Kashif Rasul, Calvin Seward, Ingmar Schuster, and Roland Vollgraf.
\newblock Autoregressive denoising diffusion models for multivariate
  probabilistic time series forecasting.
\newblock In \emph{International Conference on Machine Learning}, pages
  8857--8868. PMLR, 2021.

\bibitem[Rios et~al.(2020)Rios, van Stein, Menzel, Back, Sendhoff, and
  Wollstadt]{rios2020feature}
Thiago Rios, Bas van Stein, Stefan Menzel, Thomas Back, Bernhard Sendhoff, and
  Patricia Wollstadt.
\newblock Feature visualization for 3d point cloud autoencoders.
\newblock In \emph{2020 International Joint Conference on Neural Networks
  (IJCNN)}, pages 1--9. IEEE, 2020.

\bibitem[Rombach et~al.(2022)Rombach, Blattmann, Lorenz, Esser, and
  Ommer]{rombach2022high}
Robin Rombach, Andreas Blattmann, Dominik Lorenz, Patrick Esser, and Bj{\"o}rn
  Ommer.
\newblock High-resolution image synthesis with latent diffusion models.
\newblock In \emph{Proceedings of the IEEE/CVF Conference on Computer Vision
  and Pattern Recognition}, pages 10684--10695, 2022.

\bibitem[Satake(1982)]{satake1982fabric}
M~Satake.
\newblock Fabric tensor in granular materials.
\newblock In \emph{IUTAM-Conference on Deformation and Failure of Granular
  Materials, 1982}, pages 63--68, 1982.

\bibitem[Satake(1983)]{satake1983fundamental}
M~Satake.
\newblock Fundamental quantities in the graph approach to granular materials.
\newblock In \emph{Studies in Applied Mechanics}, volume~7, pages 9--19.
  Elsevier, 1983.

\bibitem[Sohl-Dickstein et~al.(2015)Sohl-Dickstein, Weiss, Maheswaranathan, and
  Ganguli]{sohl2015deep}
Jascha Sohl-Dickstein, Eric Weiss, Niru Maheswaranathan, and Surya Ganguli.
\newblock Deep unsupervised learning using nonequilibrium thermodynamics.
\newblock In \emph{International Conference on Machine Learning}, pages
  2256--2265. PMLR, 2015.

\bibitem[Song et~al.(2020)Song, Meng, and Ermon]{song2020denoising}
Jiaming Song, Chenlin Meng, and Stefano Ermon.
\newblock Denoising diffusion implicit models.
\newblock \emph{arXiv preprint arXiv:2010.02502}, 2020.

\bibitem[Thomas et~al.(1995)Thomas, Wiltshire, and Williams]{thomas1995use}
MC~Thomas, RJ~Wiltshire, and AT~Williams.
\newblock The use of fourier descriptors in the classification of particle
  shape.
\newblock \emph{Sedimentology}, 42\penalty0 (4):\penalty0 635--645, 1995.

\bibitem[Urbach et~al.(2020)Urbach, Ben-Shabat, and
  Lindenbaum]{urbach2020dpdist}
Dahlia Urbach, Yizhak Ben-Shabat, and Michael Lindenbaum.
\newblock Dpdist: Comparing point clouds using deep point cloud distance.
\newblock In \emph{Computer Vision--ECCV 2020: 16th European Conference,
  Glasgow, UK, August 23--28, 2020, Proceedings, Part XI 16}, pages 545--560.
  Springer, 2020.

\bibitem[Varfolomeev et~al.(2019)Varfolomeev, Yakimchuk, and
  Safonov]{varfolomeev2019application}
Igor Varfolomeev, Ivan Yakimchuk, and Ilia Safonov.
\newblock An application of deep neural networks for segmentation of
  microtomographic images of rock samples.
\newblock \emph{Computers}, 8\penalty0 (4):\penalty0 72, 2019.

\bibitem[Vlahini{\'c} et~al.(2017)Vlahini{\'c}, Kawamoto, And{\`o}, Viggiani,
  and Andrade]{vlahinic2017computed}
Ivan Vlahini{\'c}, Reid Kawamoto, Edward And{\`o}, Gioacchino Viggiani, and
  Jos{\'e}~E Andrade.
\newblock From computed tomography to mechanics of granular materials via level
  set bridge.
\newblock \emph{Acta Geotechnica}, 12:\penalty0 85--95, 2017.

\bibitem[Vlassis and Sun(2021)]{vlassis2021sobolev}
Nikolaos~N Vlassis and WaiChing Sun.
\newblock Sobolev training of thermodynamic-informed neural networks for
  interpretable elasto-plasticity models with level set hardening.
\newblock \emph{Computer Methods in Applied Mechanics and Engineering},
  377:\penalty0 113695, 2021.

\bibitem[Vlassis and Sun(2022)]{vlassis2022component}
Nikolaos~N Vlassis and WaiChing Sun.
\newblock Component-based machine learning paradigm for discovering
  rate-dependent and pressure-sensitive level-set plasticity models.
\newblock \emph{Journal of Applied Mechanics}, 89\penalty0 (2), 2022.

\bibitem[Vlassis and Sun(2023{\natexlab{a}})]{vlassis2023denoising}
Nikolaos~N Vlassis and WaiChing Sun.
\newblock Denoising diffusion algorithm for inverse design of microstructures
  with fine-tuned nonlinear material properties.
\newblock \emph{arXiv preprint arXiv:2302.12881}, 2023{\natexlab{a}}.

\bibitem[Vlassis and Sun(2023{\natexlab{b}})]{vlassis2023geometric}
Nikolaos~N Vlassis and WaiChing Sun.
\newblock Geometric learning for computational mechanics part ii: Graph
  embedding for interpretable multiscale plasticity.
\newblock \emph{Computer Methods in Applied Mechanics and Engineering},
  404:\penalty0 115768, 2023{\natexlab{b}}.

\bibitem[Vlassis et~al.(2020)Vlassis, Ma, and Sun]{vlassis2020geometric}
Nikolaos~N Vlassis, Ran Ma, and WaiChing Sun.
\newblock Geometric deep learning for computational mechanics part i:
  Anisotropic hyperelasticity.
\newblock \emph{Computer Methods in Applied Mechanics and Engineering},
  371:\penalty0 113299, 2020.

\bibitem[Wang et~al.(2019)Wang, Sun, Liu, Sarma, Bronstein, and
  Solomon]{wang2019dynamic}
Yue Wang, Yongbin Sun, Ziwei Liu, Sanjay~E Sarma, Michael~M Bronstein, and
  Justin~M Solomon.
\newblock Dynamic graph cnn for learning on point clouds.
\newblock \emph{Acm Transactions On Graphics (tog)}, 38\penalty0 (5):\penalty0
  1--12, 2019.

\bibitem[Wieghardt(1975)]{wieghardt1975experiments}
K~Wieghardt.
\newblock Experiments in granular flow.
\newblock \emph{Annual Review of Fluid Mechanics}, 7\penalty0 (1):\penalty0
  89--114, 1975.

\bibitem[Wiesmann et~al.(2021)Wiesmann, Milioto, Chen, Stachniss, and
  Behley]{wiesmann2021deep}
Louis Wiesmann, Andres Milioto, Xieyuanli Chen, Cyrill Stachniss, and Jens
  Behley.
\newblock Deep compression for dense point cloud maps.
\newblock \emph{IEEE Robotics and Automation Letters}, 6\penalty0 (2):\penalty0
  2060--2067, 2021.

\bibitem[Williams et~al.(2019)Williams, Schneider, Silva, Zorin, Bruna, and
  Panozzo]{williams2019deep}
Francis Williams, Teseo Schneider, Claudio Silva, Denis Zorin, Joan Bruna, and
  Daniele Panozzo.
\newblock Deep geometric prior for surface reconstruction.
\newblock In \emph{Proceedings of the IEEE/CVF Conference on Computer Vision
  and Pattern Recognition}, pages 10130--10139, 2019.

\bibitem[Williams et~al.(2022)Williams, Gojcic, Khamis, Zorin, Bruna, Fidler,
  and Litany]{williams2022neural}
Francis Williams, Zan Gojcic, Sameh Khamis, Denis Zorin, Joan Bruna, Sanja
  Fidler, and Or~Litany.
\newblock Neural fields as learnable kernels for 3d reconstruction.
\newblock In \emph{Proceedings of the IEEE/CVF Conference on Computer Vision
  and Pattern Recognition}, pages 18500--18510, 2022.

\bibitem[Wu et~al.(2021)Wu, Pan, Zhang, Wang, Liu, and Lin]{wu2021density}
Tong Wu, Liang Pan, Junzhe Zhang, Tai Wang, Ziwei Liu, and Dahua Lin.
\newblock Density-aware chamfer distance as a comprehensive metric for point
  cloud completion.
\newblock \emph{arXiv preprint arXiv:2111.12702}, 2021.

\bibitem[Yang et~al.(2018)Yang, Feng, Shen, and Tian]{yang2018foldingnet}
Yaoqing Yang, Chen Feng, Yiru Shen, and Dong Tian.
\newblock Foldingnet: Point cloud auto-encoder via deep grid deformation.
\newblock In \emph{Proceedings of the IEEE conference on computer vision and
  pattern recognition}, pages 206--215, 2018.

\bibitem[Zhang and Chen(2018)]{zhang2018link}
Muhan Zhang and Yixin Chen.
\newblock Link prediction based on graph neural networks.
\newblock \emph{Advances in neural information processing systems}, 31, 2018.

\bibitem[Zhang et~al.(2022)Zhang, Yin, and Chen]{zhang2022image}
Pin Zhang, Zhen-Yu Yin, and Qiushi Chen.
\newblock Image-based 3d reconstruction of granular grains via hybrid algorithm
  and level set with convolution kernel.
\newblock \emph{Journal of Geotechnical and Geoenvironmental Engineering},
  148\penalty0 (5):\penalty0 04022021, 2022.

\end{thebibliography}

\end{document}